# Free Electrons Holes and Novel Surface Polar Order in Tetragonal BaTiO$_3$ Ground States

Y. Watanabe[1,2]*, D. Matsumoto[1], Y. Urakami[1], A. Masuda[1], S. Miyauchi[1], S. Kaku[1], S.-W. Cheong[3], M. Yamato[1], and E. Carter[3]

[1]University of Hyogo, Himeji, Japan, [2]Kyushu University, Fukuoka, Japan, ATT Bell Lab, Murray Hill, NJ, USA

*watanabe@phys.kyushu-u.ac.jp

We find novel polar orders that yield electron ($e^-$) and hole ($h^+$) gas and depend on surface terminations, using density functional theory (DFT) that, unlike existing reports, relaxed all the ion positions of ATiO$_3$ having spontaneous polarization $P_S$ (A: alkali earth metal). By the experiments of atomic-oxygen cleaned surfaces of BaTiO$_3$, we find both $e^-$ and $h^+$ gas that are proven to originate from $P_S$ and constrain electrostatic potential, which has been missing. These experiments that remarkably agree with the DFTs of defect free BaTiO$_3$ reveal the properties of $P_S$-originated $e^-h^+$ and, for ferroelectric basics, an $e^-h^+$-posed intrinsic constraint on depolarization field arising from $P_S$ in proper time ranges.

Built-in fields in insulators accumulate interests for functional two-dimensional (2D) electrons ($e^-$) and holes ($h^+$) gases [1-3]. In particular, depolarization field ($E_d$) arising from spontaneous polarization ($P_S$) of insulating ferroelectrics is expected to form 2D $e^-$ and $h^+$ in novel geometries [4-8]. Theoretically, these $e^-$ and $h^+$ (abbreviated as $e^-h^+$) exist on positive $P_S$ ($P+$) and on negative $P_S$ ($P-$) faces, respectively, accompanying the potential difference between these faces $\Delta\phi$ that is close to ferroelectric bandgap $E$g (Fig. 1(a)) [6-8]. Therefore, we conjecture that the $e^-h^+$ constrains the existence-form of $E_d$ [7] that is considered as a basis underlying ferroelectric properties [9-14]. The validation of this conjecture, yet unaccepted, requires the experimental proof of intrinsicality, which is missing.

When polaronic trapping is negligible, the 2D conductance of these layers ~ $P_S\mu$ is theoretically close to the minimum 2D metallic conductance [15] ($\mu$ ~ 1 cm$^2$/Vs [16]), and the free carrier density $n$ is approximately $P_S/el_{eh}$ ~ $10^{20-21}$ cm$^{-3}$ or $P_S/e$ ~ $10^{14}$ cm$^{-2}$ ($e$: elementary charge, $l_{eh}$: thickness of the layer ~ 1nm) [8, 17-19]. Such metal-like $n$ makes Schottky-barrier unimportant [20] to yield ohmic current-voltage ($IV$) characteristics down to low field. Contrarily, nonohmic (nonlinear) conductance in insulators is due to nonequilibrium carriers injected from electrode or excited from traps [21], suggesting that the equilibrium $n$ in the conduction path is negligibly small (SM-1; SM-# denote Section# in Supplemental Material).

Such $e^-h^+$ is considered to form at $P+$ to $P+$ ($P+P+$) and $P-$ to $P-$ ($P-P-$) domain boundaries (DB) and ferroelectric surfaces and interfaces (Fig. S1; Fig. S# is in Supplemental Material) [8,18-19, 22-28], where enhanced conductance at these DBs is reported as evidence. However, these DBs show either $e^-$ or $h^+$, non-ohmicity, no enhanced conductance at low field, and low $n$ (7500 cm$^{-2}$ at high field [22]. Some semiconducting ferroelectrics show ohmicity with $n < 10^{16}$ cm$^{-3}$ at high field but only $h^+$ conduction modestly larger at DBs than at bulk parts [23,24]). $\Delta\phi$ ~ $E$g is not reported. These characteristics disagree with the aforementioned theoretical expectations of defect-free ferroelectrics [6-8, 17-19]. Further, these measurements are conducted with field sufficiently high to induce resistance switching [29-31] and the migration of the ions/defects [32] that become $e^-h^+$-suppliers and closely resemble resistance switching [18]. These experimental properties might be explained by small polarons [18,21] but are consistent with defect-$P_S$ complexes, i.e. defects at these DBs proposed by studies including electron microscopy and density functional theory (DFT) [33-37]. This interpretation is fortified by the defect-induced drastic reduction in $n$ and $E_d$ and disappearance of $h^+$ (Fig. S1(b)).

Here, by resolving these issues, experiments and DFTs with/without electron correlation (and hybrid functional) for polaron reveal the $P_S$-originated $e^-h^+$ and non-contact controls of $P_S$ as a functional field effect. The DFTs that unlike the literature relaxed all the ion positions of ferroelectric-phase disclose polar orders determined by BaO- and $TiO_2$-terminations, which explains experimental $n$ and the location of $e^-$ that are different those of $h^+$.

The electrostatics of the polarization charge ($-\Delta P$) at $P+P+$ and $P-P-$ DBs, surfaces, and insulator/ferroelectric is virtually identical [38]. While $P_S$ terminates stably at surfaces even without defects, these DBs are unstable without defects and, hence, exist at defects [31-37]. Therefore, we study the free surfaces with minimal defects of $BaTiO_3$ single crystals (SM-2(Methods)). Further, we applied only low external fields without contacting the surfaces in order to eliminate extrinsic conductance and contamination, e.g. from the cantilever of the scanning probe microscope (SPM) [32].

Atomically flat $BaTiO_3$ surfaces were electroded (Fig. 1(b)). To keep clean surfaces, all the measurements were done in an ultrahigh vacuum (UHV) in darkness. Because conventional cleaning of UHV-heating (> 1100 K) creates countless oxygen vacancies (O-vacancies), sample surface is cleaned at 300 K by atomic O of which the oxidation capability is equivalent to $10^{10}$ atm $O_2$ (Fig. 1(c)) [39] and an electrostatic method (Fig. S2). The stable outermost layer of the [001] [100] surfaces of $ATiO_3$ (A: Ba, Sr) is AO in an oxidizing atmosphere and $TiO_2$ in a reducing atmosphere and air [40-42], and buckling of the $TiO_2$ is reported for paraelectrics [42,43]. The $P_S$ in the gap between two electrodes, T1 and T2, was oriented by applying 80 V~160 V/mm between T1 (T2) and the bottom electrode B (Figs. 1(d) and S3). This field was the maximum applied field in the experiments and not on the surface conduction path. Such low applied field in the gap and the large width of the gap (0.012~0.3 mm) prohibited migration of ions and defects into the main part of the gap. That is, the $P_S$-orientation was controlled without contacting the target surface, thereby eliminating conductance unrelated with $P_S$. The poled states as shown in Fig. 1 were obtained by performing poling while cooling the sample from $T_C$ (~ 400 K) to 300 K, which is a slightly reducing condition, and the surface-terminations of $BaTiO_3$ in Fig. 1 is considered as $TiO_2$ [40-42].

The orientation of $P_S$ (// $c$-axis) was monitored by polarization microscopy, piezoelectric response microscopy (PFM), and the nonlinear capacitance (Fig. S4). To retain cleanness, the surface potential and topography were measured in true non-contact mode that uses Van der Waals force. To avoid contacting the surface, the domains were mainly identified from the potential images perfectly agreeing with the PFM images (Fig. S5). Capacitance $C$ and $IV$ curves for determining the conductance $\sigma$ between T1 and T2 ($\sigma_{T-T}$, current $I_{T-T}$, Fig. 1(e)), T1 and B ($\sigma_{T-B}$, current $I_{T-B}$), and T2 and B ($\sigma_{T-B}$, $I_{T-B}$) were acquired; here, $\sigma_{T-B}$ is the bulk conductance $\sigma_{bulk}$ (Fig. S6). Carrier type identification was supplemented by chemistry [44,45] (Fig. S7). $T_C$ was detected using polarization microscopy and from the $T$-dependence of $C$. The maximum applied voltage and the field for the $IV$ measurements were 0.5 V and 1 ~ 40 mV/μm, respectively to avoid carrier injection and resistance switching. Because the BaTiO$_3$ was $10^5$ times thicker than the thickness of $e^-h^+$ layers [8,17], $\sigma_{T-T}$ due to bulk appeared at high $T$ (Fig. 1(e), SM-3).

Figures 1(f)-1(j) show the polarization microscopy images, $IV$ hysteresis curves of each state, and corresponding $T$-dependences of $\sigma_{T-T}$ and $P_S$ (by Ginzburg-Landau theory). The initial state was unpoled and $ac$ domained; the $P-$ and $P+$ states formed subsequently and were measured. After an additional $P+$ poling, the $IV$ curves and $\sigma_{T1-T2}(P++)$ were measured (Figs. 1(g) and 1(j)). The ohmicity down to low fields proves that conductance was due to the equilibrium carriers. $\sigma_{T-T}$ of the unpoled states ($\sigma_{T-T}^{unpoled}$) was the same in UHV and air wherein free $e^-h^+$ at surfaces disappeared by adsorbates (Fig. S8). Therefore, $\sigma_{T-T}^{unpoled}$ was due to the thick bulk, and $\sigma_{surface} = \sigma_{T-T} - \sigma_{T-T}^{unpoled}$ (This means also $\sigma_{surface}^{unpoled} = 0$), where $\sigma_{surface}$ denotes the surface conductance. Alternatively, $\sigma_{surface} \approx \sigma_{T-T} - \sigma_{T-B}$, because the bulk conductance $\sigma_{T-B}$ approximately equaled $\sigma_{T-T}^{unpoled}$ (Figs. 1(h) and 1(j)). Therefore, below $T^*$ of Fig. 1(j), $\sigma_{T-T}$ of all of the poled states was $\sigma_{surface}$, which applies also to $I_{T-T}$.

$\sigma_{surface}^{unpoled} = 0$ and the disappearance of $\sigma_{surface}$ (= $\sigma_{T-T} - \sigma_{T-T}^{unpoled}$) above $T_C$ that coincided with that of $P_S$ prove that $\sigma_{surface}$ was induced by $P_S$ (Fig. 1(i)); the approach of $\sigma_{T-T}$ of the $P++$ state to $\sigma_{T-T}^{unpoled}$ was insufficient because we did not wait for the slow approach. Furthermore, in the repeated formation of different states, only $P+$, $P-$, or unpoled determined $\sigma_{T-T}$, whereas the number of vacuum heatings to 410 K, which would have increased O-vacancy, did not affect $\sigma_{T-T}$ (Fig. S9, SM-4, SM-5).

The $IV$ hysteresis curve of the $P+$ surface after a long-time poling exhibited ohmic conductance with a conductivity of 0.1 $\Omega^{-1}$cm$^{-1}$ for $l_{eh} \sim 2$ nm (Fig. 2(a)) and persisted for at least 15 h (Fig. 2(b)). This conductivity is 1/80th the theoretical conductivity of a continuous atomically flat surface for $\mu = 1$ cm$^2$/Vs [16] and close to the minimum metallic conductance [15]. Subsequently, the BaTiO$_3$ was exposed to N$_2$ gas, whereby the adsorbates in the N$_2$ were expected to diminish $E_d$, produce traps of

$e^-h^+$, and thereby reduce $\sigma_{surface}$. After the exposure, $\sigma_{T-T}$ decreased to 1/500th its value before the exposure (Figs. 2(b) and S10). However, it was still 100 times higher than $\sigma_{T-B}$ ($\approx \sigma_{T-T}^{unpoled}$). Because $\sigma_{surface} \approx \sigma_{T-T} - \sigma_{T-B}$, this shows that the surface conduction layer survived after the exposure and demonstrates its robustness.

Figure 3 shows potential $\phi$ and $\sigma_{T-T}$ of each poled state of another $BaTiO_3$, where no measurements of the $T$-dependence were conducted to minimize the formation of O-vacancy. $BaTiO_3$ in Figs. 3(a)-3(e) experienced no vacuum heating, while $BaTiO_3$ in Figs. 3(f)-3(j) experienced vacuum heating to 410 K. The initial state directly after O irradiation (Fig. 3(a)) was a multi-domained one consisting of $P+$ and $P-$ domains; the green areas on the left of the white dashed lines in the images are grounded electrodes. All the experiments were performed in non-contact except for PFM at 4 points.

First, we will examine Figs. 3(f)-3(j), which depict properties after vacuum heating and hence are probably of the $TiO_2$ surface as in Fig. 1. By positive and negative poling at 300 K, $\phi$ of the free surface immediately became uniformly positive and negative with respect to the electrode, respectively (Figs. 3(g)-3(i)), and $\sigma_{surface}$ appeared. These surfaces of positive and negative $\phi$ were $P+$ and $P-$ surfaces, respectively (Fig. S5), which was confirmed by PFM at the four points of Fig. 3(h). Therefore, the ohmic $IV$ curves in blue and red (Fig. 3(j)) are of $P-$ and $P+$ surfaces, respectively, and closely resemble the $IV$ curves in Fig. 1(h) except for the conductance lower than those in Fig. 1(h) (because of the inferior surface flatness).

$|\phi|$ in Figs. 3(g)-3(i) was 100 times lower than the unscreened $|\phi|$ by Kittel theory [9] that yields a lower bound on $|\phi|$ for vortex domains [10] (Fig. S11, SM-6). This means that negative (positive) charges moved to the $P+$ ($P-$) surface. These charges had high mobility because the screening occurred within milliseconds (Fig. 3(f)). The difference in $e\phi$ between the $P+$ and $P-$ surfaces, $e\Delta\phi$, was close to $Eg$ of $BaTiO_3$ (Fig. 3(i)). $\sigma_{T-T}$ of the $P+$ and $P-$ states became higher down to 0 V than $\sigma_{T-T}^{unpoled}$ (Fig. 3(j)), showing the emergence of $\sigma_{surface}$ (This $\sigma_{T-T}^{unpoled}$ was estimated from $\sigma_{T-T}^{unpoled}$ of Fig. 3(e), because $\sigma_{T-T}^{unpoled}$ was independent of vacuum heating (Fig. S9)). Similar poling repeatedly formed $P+$ and $P-$ surfaces, their $\sigma_{surface}$'s correlated completely with $\phi$ (Figs. 3(i) and 3(j)), and the potential profiles obeyed those estimated by $e^-h^+$ depletion theory (Figs. 3(i) and S12, SM-7). These observations consistently show that $e^-$ and $h^+$ appeared on $P+$ and $P-$ surfaces, respectively, through $E_d$ (Fig. 1(a)).

The properties of this sample before experiencing any vacuum heating also showed that high-mobility negative (positive) charges moved to the $P+$ ($P-$) surface and induced $\sigma_{surface}$, and $e\Delta\phi \approx E_g$ (Figs. 3(b)-3(e)). However, some of these properties were different from those after heating. In Fig. 3(e), $\sigma_{surface}$ of $P-$ ($h^+$) was higher than $\sigma_{surface}$ of $P+$ ($e^-$); the vacuum heating reduced $\sigma_{surface}$ of $h^+$ and enhanced $\sigma_{surface}$ of $e^-$. This unusual characteristic fortifies the present potential-based identification of carrier types because, in $ATiO_3$, $h^+$ conductance increases by over-oxygenation and decreases by reduction in O, while $e^-$ conductance increases by reduction in O [44,45]. Moreover, the increase in the potential of the free surface $-e\phi$ caused by heating (Figs. 3(d) and 3(i)) matched the theoretical increase (Figs. 4(a) and 4(c), SM-8).

The experimental results in Figs. 1-3 match DFT calculations of fully-relaxed tetragonal $BaTiO_3$ without defects in vacuum, but not calculations including O-vacancy. We examine $TiO_2$- and BaO-terminated surfaces (Figs. 4(a)-4(d)), which are stable in an ordinary and reducing atmosphere and an oxidizing atmosphere, respectively [40-42].

DFT of $TiO_2$-terminated $BaTiO_3$ including electron correlation for polaron [46,47] agreed with the results obtained after the $BaTiO_3$ was heated. The theoretical potential difference $e\Delta\phi \approx E_g$ between the $P+$ and $P-$ surface (Fig. 4(a)) matched the experimental value (Fig. 3(i), SM-9). The density of states (DOS) of the $TiO_2$ layer indicates there is a metallic $Ti_{3d}$-like $e^-$ layer near the $P+$ surface and a small density of $O_{2p}$-like $h^+$ at the $P-$ surface (Fig. 4(b)), which explains $\sigma_{surface}$ of the $P+$ surface much higher than $\sigma_{surface}$ of the $P-$ surface in Figs. 1 and 3(j). This enhancement of $\sigma_{surface}$ and the robustness against adsorbates in Fig. 2(b) originates from the intrinsic $P+P+$ polar order formed by the buckling of topmost $TiO_2$ at the $P+$ surface that enhances polarization charge density $-\Delta P$ (inset of Fig. 4(b); this order exists also with O-vacancy (Fig. S13)). DFT of BaO-terminated $BaTiO_3$ (Figs. 4(c) and 4(d)) agreed with the results of the $BaTiO_3$ experiencing no vacuum heating (Figs. 3(b) - 3(e)); the DOS showed that the density of $h^+$ near the $P-$ surface was higher than that of $e^-$ at the $P+$ surface, and the experimental $\sigma_{surface}$ of the $P-$ surface was higher than $\sigma_{surface}$ of the $P+$ surface (Fig. 3(e)). This enhancement in $h^+$ density originated the intrinsic $P-P-$ polar order formed by the buckling of outermost $TiO_2$ at the $P-$ surface that enhances $-\Delta P$ (inset of Fig. 4(d)).

These agreements between DFT and experiment fortifies attribution of the surface before and after heating as BaO- and TiO$_2$-terminated surfaces, respectively. Such attribution is consistent with the drastic changes induced by the first heating and the negligible changes induced by subsequent heatings (Fig. 3) because TiO$_2$-terminated surfaces are stable against reduction in O [40-42]. The properties of BaTiO$_3$ surface, particularly $e^-$ state different from $h^+$ state, appear to be general ones for ATiO$_3$-type ferroelectric surface, as indicated by the analogous results for SrTiO$_3$ (Fig. S14). Contrastingly, DFT considering O-vacancy showed that O-vacancy extinguishes $E_d$ (i.e. $e\Delta\phi \ll E_g$) and deletes DOS at $E_F$ at the $P-$ surface (Figs. 4(e) and 4(f)), which coincided with the DFT of DB with O-vacancy (Fig. S1(b)) but did not match our experiments.

In summary, we attained clean surfaces of BaTiO$_3$ with so few O-vacancies to have BaO-terminations. We reversibly formed $P+$ and $P-$ free surfaces without contacting the surface, which was enabled by $P_S$-induced $e^-h^+$ as intrinsic electrodes and allowed avoiding ion migration and charge injection. These experiments revealed high ohmic conductance down to mV/μm order with on/off ratio up to $10^5$ that disappeared above $T_C$ and increased with surface flatness and the reduction of defects. The measurements of potential $\phi$ uncovered high-speed charges of negative sign on the $P+$ surface and of positive sign on $P-$ surface that induced conductance and $e\Delta\phi \sim E_g$ between these surfaces. These characteristics unambiguously point to there being a free $e^-$ layer on the $P+$ surface and a free $h^+$ layer on the $P-$ surface that arise from $P_S$. The comparison between the experimental results and DFT shows that these characteristics are of defect-free BaTiO$_3$. Further, DFT of thick BaTiO$_3$ predicted novel polar orders: $P+P+$ at TiO$_2$-surface and $P-P-$ at BaO-surface. This was experimentally supported by $h^+$ conductance higher than $e^-$ conductance before heating, $h^+$ conductance much lower than $e^-$ conductance after heating, surface potential, and inward location of $e^-$ layer suggested by Fig. 2. Because of the persistence and the generality of the present $e^-h^+$ as shown by DFT on SrTiO$_3$ and DB, $e^-$ and $h^+$ layers are essential for defect-free ferroelectrics as an intrinsic constraint on $E_d$ ($|e\Delta\phi| < E_g$).

Because metal-like $e^-h^+$ formed by $P_S$ was experimentally shown and DFT considering electron correlation showed metallic states, the reported properties of conducting DBs: non-ohmicity and little conductance at low field [18-19, 22-27] were due not to polaron but to defects [33-36], i.e. defect-$P_S$ complex; these DBs are regarded as excellent facilitators of nonohmic conduction and resistance switching through enhanced ion-mobility [37] and the reduction of work-function (Fig. S1(b)). Such screening by defects and adsorbates that render $E_d$ almost negligible (Figs. 2(b), 4(e), and S1(e)) is usually ignored [11-13] despite inevitable existence in thin films [1-3]. By clarifying the properties of the DBs [18-19, 22-27, 33-36], these studies proved that defect-screening must be considered in experiments. This suggests a hypothesis that $E_d$ is unscreened in a short time-scale and screened by defects/adsorbates in a long time-scale, where $e^-h^+$ constructs the road from the former to the latter (SM-10).

This work was supported by Murata Science Foundation and JSPS KAKENHI Grant No. JP19K21853. The discussions with Daniel C. Tsui of Princeton Univ., J. Georg Bednorz of IBM, Peter Blöchel of Clausthal Univ. Tech., Gordon. A. Thomas of ATT Bell and NJIT, M. Arai of NIMS, S. A. Lyon of Princeton Univ. are greatly acknowledged.

Figure captions

FIG. 1 Non-contact control of polarization and conductance of free surface. (a) Band under $P_S$ forming $e^-h^+$. (b) Electrodes: T1, T2, and B. (c) Cleaning by atomic oxygen. (d) Forming a $P+$ state. (e) Bulk contribution to $\sigma_{T-T}$. (f) Polarization microscopy of $P+$, $P-$ and unpoled state ($P\rightarrow$). Electrodes and surfaces with $P+$ or $P-$ are dark. (g)(h) *IV* hysteresis loops of (g) $P++$ and (h) the states of (g). (i) *T*-dependence of $P_S$. (j) *T*-dependence of $\sigma_{T-T}$ and $\sigma_{T-B}$ measured simultaneously with $\sigma_{T-T}$ of $P\rightarrow$. Filled and open symbols represent $\sigma_{T-T}$ ($I_{T-T}$) and $\sigma_{T-B}$ ($I_{T-B}$), respectively.

FIG. 2 Persistence of surface conductance. (a) *IV* hysteresis loops of unpoled and $I_{T-T}$ and $I_{T-B}$ of $P+$ state (each loop is overlapped). (b) $\sigma_{T-T}$ ($\approx \sigma_{surface}$) and $\sigma_{T-B}$ ($\sim \sigma_{T-T}^{unpoled}$) before and after $N_2$ exposure.

FIG. 3 Surface potential and conductance of a same location of a free surface (a)-(e) before and (f)-(j) after heating. Surface-terminations were BaO in (a)-(e) and $TiO_2$ in (f)-(j). Chronological order: (a)-(c), (g), (h). (f) Change of $\phi$ during poling. (a)(b)(c)(g)(i) Surface potential images. Electrodes (left of white dashed lines) were fixed at 0 V. Light-blue points show PFM measurement spots. (d)(h) Cross-section of surface potential images in $-e\phi$. (e)(i) *IV* curves. In (d)(e)(i)(j), the same color corresponds to the same state, the letters a-c, g, and h show corresponding potential images, and $P+$ etc. show $P_S$ orientations in a chronological order. Green lines in (d) and (i) show depletion layer theory.

Fig. 4 $BaTiO_3$ having (c)(d) BaO- and (a)(b)(e)(f) $TiO_2$-terminated surfaces without/with O-vacancy. (e)(f) an O-vacancy at $P-$ surface. (a)(c)(e) Potential along *c*-axis direction. (b)(d)(f) Local density of states of $TiO_2$ layer along *c*-axis. Illustrations: ion positions near the top and bottom surface along *c*-axis. Green, blue, and red circles show Ba, Ti and O ions, respectively. Tripods show the orientation of unitcell dipole.

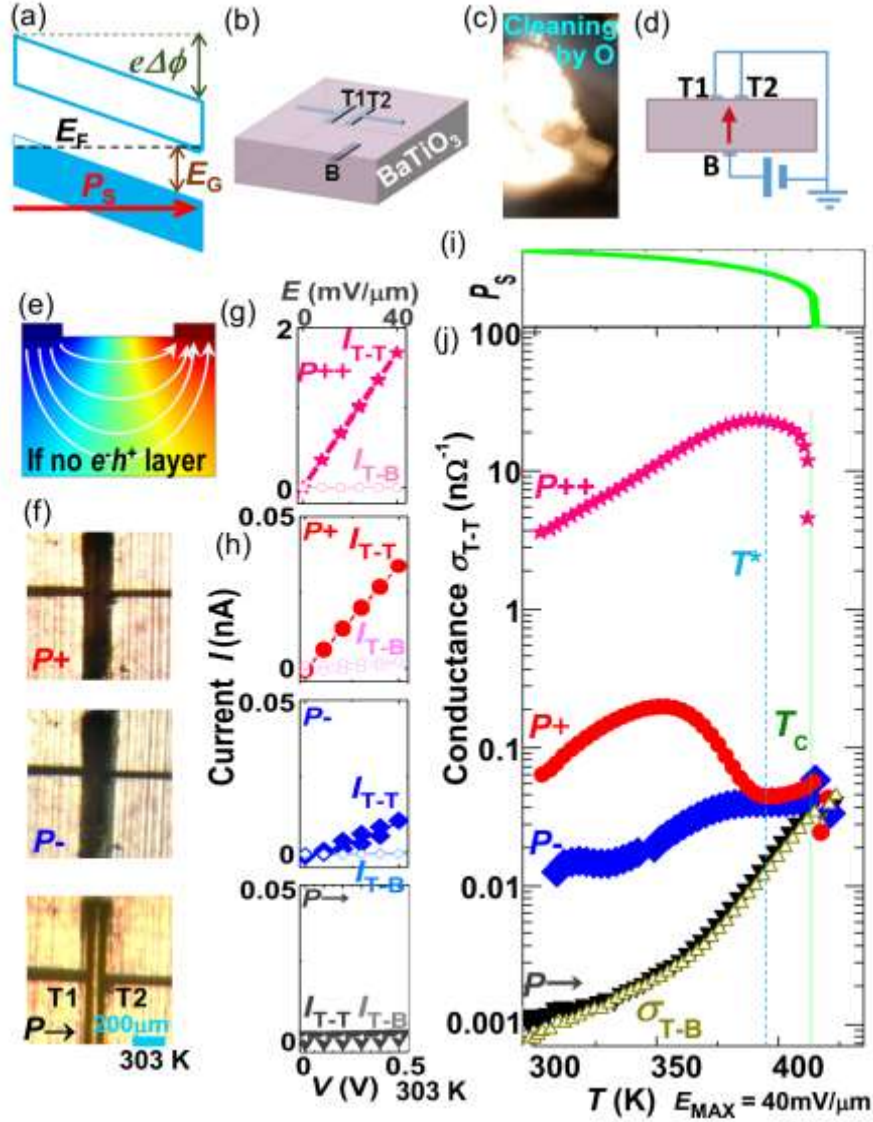

FIG. 1 Non-contact control of polarization and conductance of free surface. (a) Band under $P_S$ forming $e^-h^+$. (b) Electrodes: T1, T2, and B. (c) Cleaning by atomic oxygen. (d) Forming a $P+$ state. (e) Bulk contribution to $\sigma_{T-T}$. (f) Polarization microscopy of $P+$, $P-$ and unpoled state ($P\rightarrow$). Electrodes and surfaces with $P+$ or $P-$ are dark. (g)(h) $IV$ hysteresis loops of (g) $P++$ and (h) the states of (g). (i) $T$-dependence of $P_S$. (j) $T$-dependence of $\sigma_{T-T}$ and $\sigma_{T-B}$ measured simultaneously with $\sigma_{T-T}$ of $P\rightarrow$. Filled and open symbols represent $\sigma_{T-T}$ ($I_{T-T}$) and $\sigma_{T-B}$ ($I_{T-B}$), respectively.

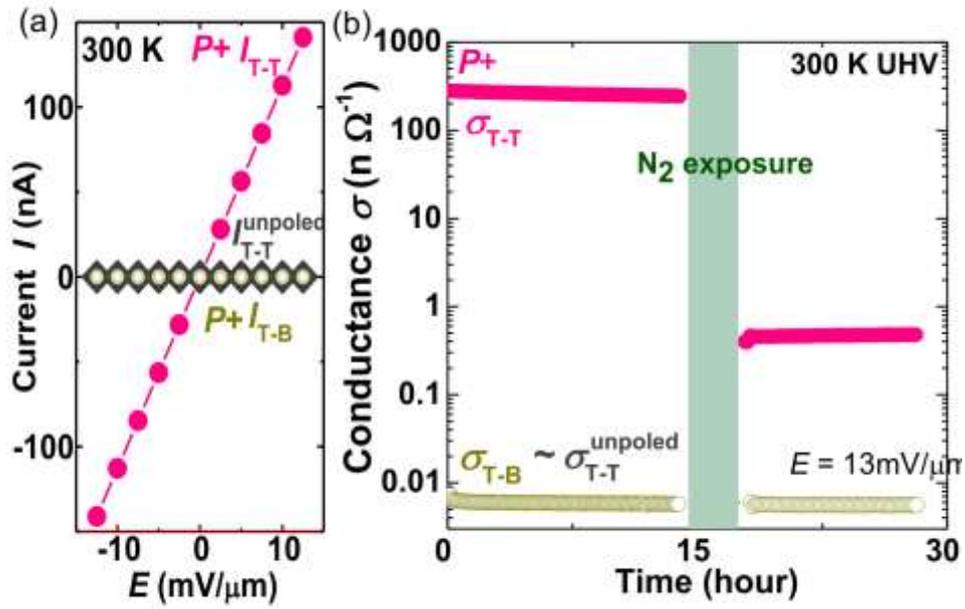

FIG. 2 Persistence of surface conductance. (a) $IV$ hysteresis loops of unpoled and $I_{T-T}$ and $I_{T-B}$ of $P+$ state (each loop is overlapped). (b) $\sigma_{T-T}$ ($\approx \sigma_{surface}$) and $\sigma_{T-B}$ ($\sim \sigma_{T-T}^{unpoled}$) before and after $N_2$ exposure

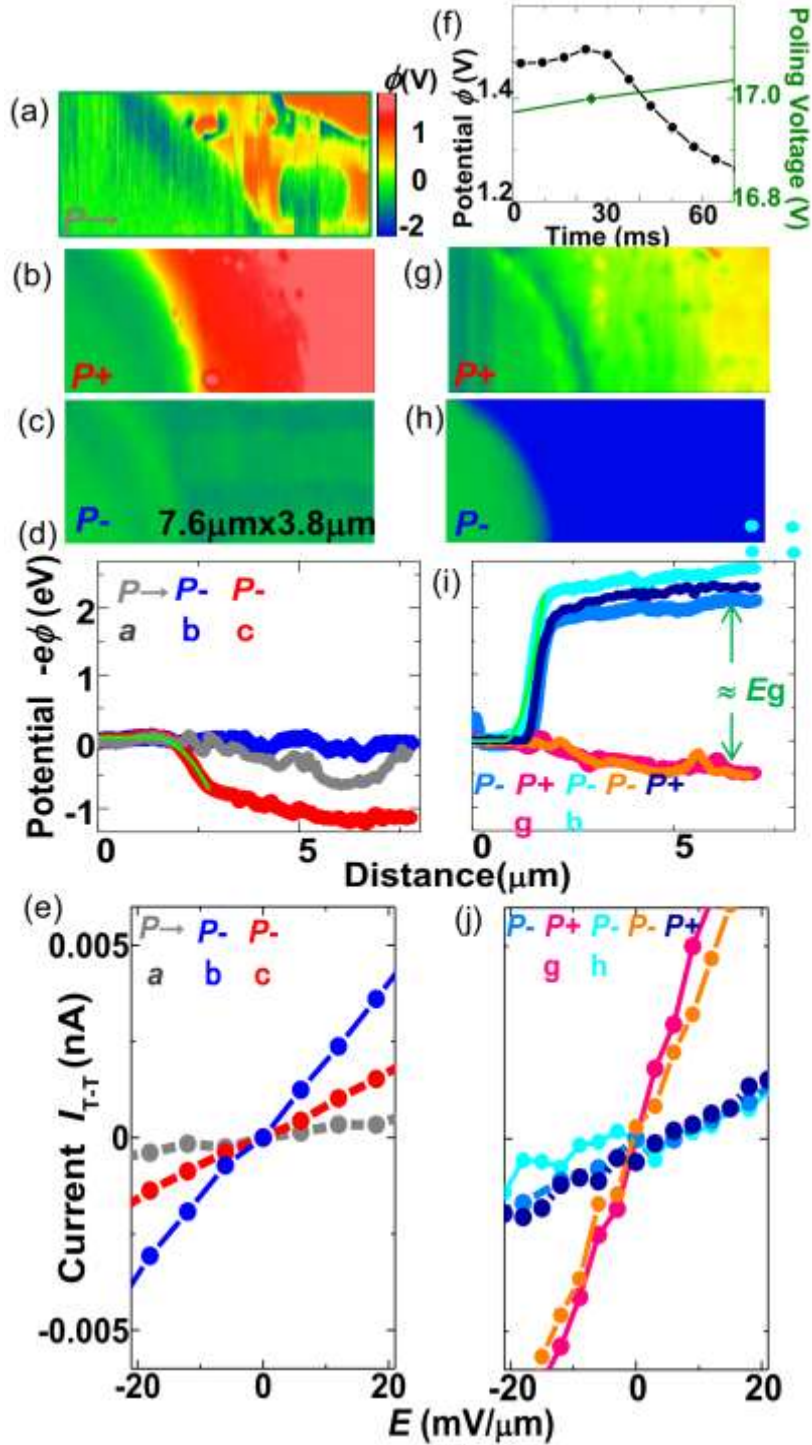

FIG. 3 Surface potential and conductance of a same location of a free surface (a)-(e) before and (f)-(j) after heating. Surface-terminations were BaO in (a)-(e) and $TiO_2$ in (f)-(j). Chronological order: (a)-(c), (g), (h). (f) Change of $\phi$ during poling. (a)(b)(c)(g)(i) Surface potential images. Electrodes (left of white dashed lines) were fixed at 0 V. Light-blue points show PFM measurement spots. (d)(h) Cross-section of surface potential images in $-e\phi$. (e)(i) IV curves. In (d)(e)(i)(j), the same color corresponds to the same state, the letters a-c, g, and h show corresponding potential images, and P+ etc. show $P_S$ orientations in a chronological order. Green lines in (d) and (i) show depletion layer theory.

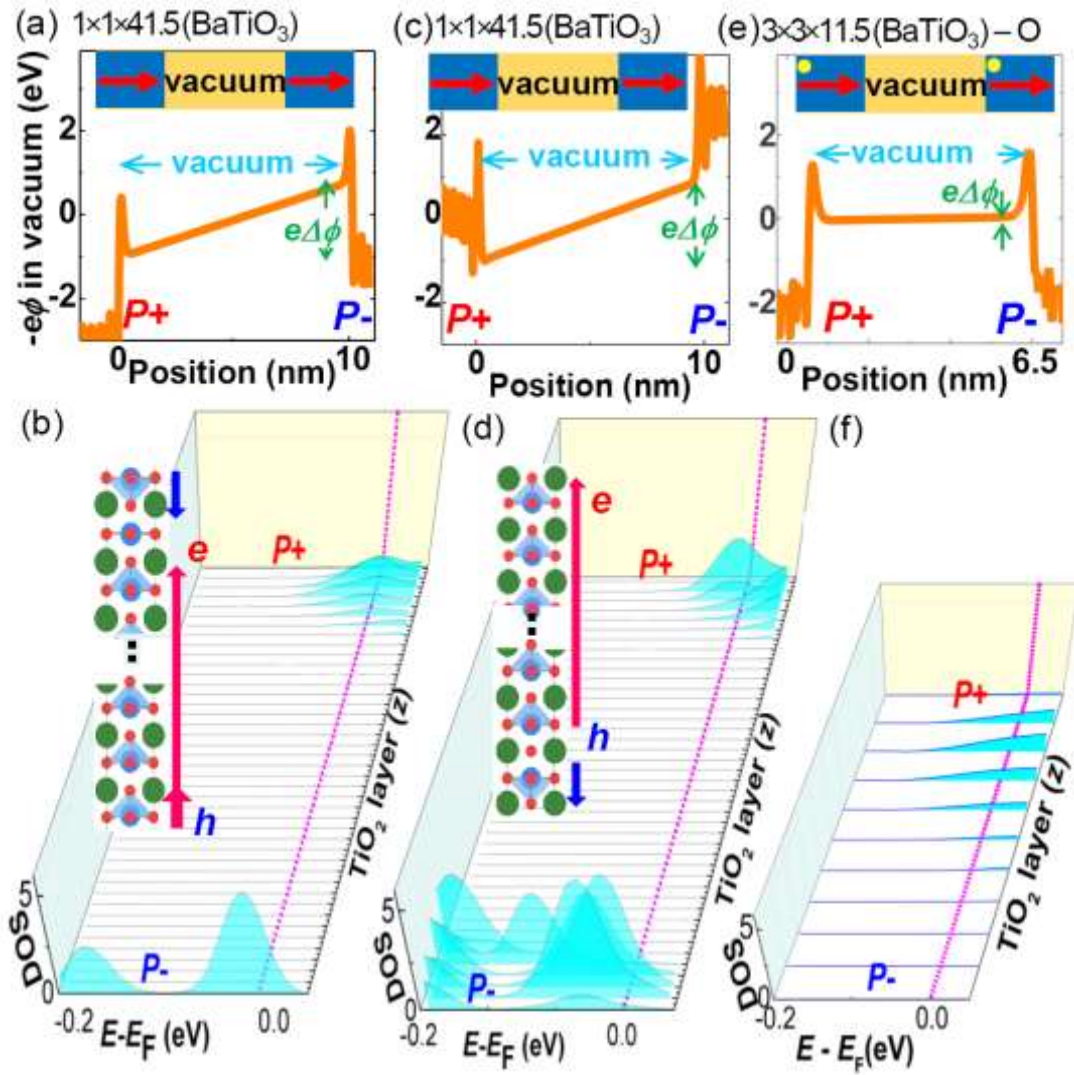

Fig. 4 BaTiO$_3$ having (c)(d) BaO- and (a)(b)(e)(f) TiO$_2$-terminated surfaces without/with O-vacancy. (e)(f) an O-vacancy at *P*- surface. (a)(c)(e) Potential along *c*-axis direction. (b)(d)(f) Local density of states of TiO$_2$ layer along *c*-axis. Illustrations: ion positions near the top and bottom surface along *c*-axis. Green, blue, and red circles show Ba, Ti and O ions, respectively. Tripods show the orientation of unitcell dipole.

# Supplemental Material for

# Free Electrons Holes and Novel Surface Polar Order in Tetragonal BaTiO$_3$ Ground States"

Y. Watanabe[1,2]*, D. Matsumoto[1], Y. Urakami[1], A. Masuda[1], S. Miyauchi[1], S. Kaku[1], S.-W. Cheong[3], M. Yamato[1], and E. Carter[3]

[1]University of Hyogo, Himeji, Japan, [2]Kyushu University, Fukuoka, Japan, [3]ATT Bell Lab, Murray Hill, NJ, USA

*watanabe@phys.kyushu-u.ac.jp

## Table of Contents

**Supplemental Material Notes**

**SM-1 Enhanced conductance alone does not mean $e^-h^+$ layer: Ohmicity at low $V$ needed.**

Conduction through a conduction path is due to three components: preexisting or equilibrium free $e^-h^+$ in the path, $e^-h^+$ excited from trapped $e^-h^+$ in the path, and injected $e^-h^+$ from electrodes [21]. When free $e^-h^+$ is abundant in the path, the first component is dominant and ohmic conduction is observed. When free $e^-h^+$ are few or trapped in path as in insulators with defects or depletion layers, the other components are dominant. In this case, non-ohmic conduction such as space-charge-limited conduction (SCLC), Poole-Frenkel, and Schottky conduction is observed. These conductions can be expressed by $ne\mu \propto E^m$ (m ≥ 0) by defining field ($E$) dependent $n(E)$ and $\mu(E)$, where an exponential dependence is approximately expressed by m = ∞. For ohmic conduction (m = 0), $n > 0$ and $\mu > 0$ at $E = 0$, meaning that free $e^-h^+$ are present in equilibrium. For nonohmic conduction (m > 0), $n = 0$ or $\mu = 0$ at $E = 0$, meaning that free $e^-h^+$ are absent in equilibrium.

In particular, Schottky conduction is due to the depletion of free $e^-h^+$ beneath the metal. But, when the free $e^-h^+$ concentration in the path is originally high, Schottky barrier becomes so thin to allow the tunneling through the barrier [20] and hence ohmic conduction. Theoretical results [6-8,17-19] and Fig. S1(a) show metallic $e^-h^+$ by $P_S$ at defect-free $P+P+$ and $P-P-$ DB. Therefore, the conduction through these $e^-h^+$ layers formed by $P_S$ should exhibit ohmic conductance extremely higher than the bulk part, when polaronic trapping is negligible.

Interestingly, Tselev [20] showed microwave conductance at DBs together with nonohmic $dc$ conductance at the DBs. Because weakly bounded $e^-h^+$ can contribute to microwave conductance, we consider this microwave conductance as evidence of abundant trapped $e^-h^+$. This result and the $IV$ characteristics agreeing with Poole-Frenkel conduction are explained by defect-$P_S$ complex or trapping of $e^-h^+$ by small polaron [18, 21]. We think that defect-$P_S$ complex is the origin of these observations, because of following reasons. When defects are sufficiently suppressed, the properties that are theoretically expected for free $e^-h^+$ appeared (Figs. 1-3). No remarkable difference between DFT with and without electron correlation for polaron was found (Polaronic states was not found).

In the view of defect-$P_S$ complex, the enhancement of conductance at DB's showing nonohmic conductance [18-19, 22-23, 25-27] is due to the suppression of Schottky barrier, Poole-Frenkel activation energy, or trap-free SCLC threshold. The $IV$ characteristics attributed to the carrier-type inversion [18] can be explained as resistance switching or field–induced defect displacements [29-32] that show similar $IV$ characteristics [29,31], because the mobility of defects increases at DBs [37] and the applied field was high [18]. In the present study, ohmic conductance or linear $I^{\text{ave}}$–$V$ was always confirmed.

**SM-2 Methods**

**Materials selection.** To realize such $e^-h^+$ intrinsically due to $P_\text{S}$, single crystals of high stoichiometry, low defects density, and an appropriate $E_\text{g}$ are necessary. Such crystal is difficult to find in ferroelectric materials, as evidenced in insufficient resistivity of many ferroelectric materials that are composed of volatile metal elements such as Pb, Li, Bi, and Mn. Pb(Ti,Zr)O$_3$ (PZT), for example, become Pb-deficient even by annealing at ~600 ºC in 1 atm air, and PbTiO$_3$, BiFeO$_3$, and Bi$_4$Ti$_3$O$_{12}$ form easily O-vacancies in vacuum heating as evidenced in the reported low resistivity [48,49]. Contrarily, color, resistivity, and dielectric properties of SrTiO$_3$ and BaTiO$_3$ did not degrade even after heating in air even at ~1000 ºC.

To prove the disappearance of $e^-h^+$ above $T_\text{C}$, $T_\text{C}$ needs to be < 420 K to avoid degradation during experiments in vacuum. BaTiO$_3$ possesses the lowest $T_\text{C}$ (> 300 K) among standard ferroelectrics [50,51], and the good stoichiometry of single crystals is shown by high insulativity (Fig. S15) and colorless transparency. Additionally, etch-pit density and etching speed indicated O-vacancies far less in single crystals than in epitaxial thin films. Hence, we used BaTiO$_3$ single crystals, while defects are massive in crystalline oxide films [2,48,49], as evidenced by the increase in resistivity following air or O$_2$ annealing [2] and very fast etching rate by acid.

**Samples.** For these reason, we used BaTiO$_3$, which is tetragonal near 300 K. BaTiO$_3$ single crystals in the present experiments (Fig. S15) were commercial top-seed-solvent-grown (TSSG) crystals from (Monotech Co. Japan and Castech Co. China) and crystals cured from ~3000 BaTiO$_3$ KF-flux grown by Carter (ATT Bell). BaTiO$_3$ in Figs 1-3 were TSSG crystals, because they are purer than KF-flux crystals. KF-flux BaTiO$_3$ crystals, which was used to cross-check the results of TSSG BaTiO$_3$ crystals, had natural atomically flat surfaces in as-grown states and experienced no mechanical polishing and exhibited excellent properties (Figs. S15(a)-15(c)). Some flux crystals were also H$_3$PO$_4$-etched to remove possible surface layers. When flux crystals are used, it is written in caption.

The purity of TSSG $BaTiO_3$ crystals was <10 ppm according to the manufacturers, which was consistent with optical density of states $N_t$ of total defects and impurity (~ $4\times10^{17}/cm^3$). Because mechanical polishing creates dislocations and artificial layers at surfaces, TSSG crystals except for those in Figs. S7 and S10(b) were $H_3PO_4$-etched and air-annealed repeatedly to etch out 100 μm and exhibited atomically flat surface. After repeated etching, the TSSG crystals exhibited nearly atomically flat surfaces with step bunching, sharp $T_C$, excellent dielectric properties, and high insulativity of 5 $p\Omega^{-1}cm^{-1}$ at 300K (Figs. S15(d)-15(f)). These properties, straight parallel natural $a/c$ domains penetrating from top to bottom surface (Fig. 1(f)), high transparency, and low coercive field of 1.5 ~ 2 kV/cm for switching were consistent with high chemical purity, stoichiometry and low defect density.

$BaTiO_3$ with rough surface showed low $\sigma_{surface}$ supporting that $\sigma_{surface}$ occurred at surface. For consistency, the results in the main text and main figures are those of two TSSG crystals: one for Figs. 1, 2, S4, S6, S9(a), S9(b), S10(a), S15(d), and S15(e) and the other (with less flat surfaces) for Figs. 3 and S12. The gap between T1 and T2 was 12.5μm×900μm for Fig. 1 and 40μm×1050μm for Fig. 2, while the width of the electrodes on a top surface was typically 30μm and that on a bottom was the same as the gap width (12.5~40μm).

Despites differences in growth methods and preparations, the major characteristics of 8 crystals having nearly atomically flat surfaces were the same. All the crystals were annealed in flowing 99.9999 % $O_2$ of 1 atm at 423K~673K before cleaning in UHV. When the surface was sufficiently flat, the conductance of TSSG surface was higher than that of as-grown KF-flux crystals with atomically flat surfaces, which we think due to the high purity of TSSG.

**New method of cleaning avoiding formation of oxygen vacancy.** The standard surface cleaning of oxide, which is 900K~1200K rapid heating in UHV, creates massive oxygen vacancies (O-vacancy). To avoid this problem, we cleaned the surface by a low-energy supersonic O beam for ~6 hours that was produced by electron cyclotron resonance ECR (Arios EMRS-211Q) from cryogenically purified $O_2$ of was initial purity 99.9999%. Emission spectroscopy identifies that all the emissions were assigned to atomic oxygen (Very small peaks of excited $O_2$ was detected, but no emissions from $H_2$, OH and $H_2O$ were detected). The cleanness by similar process is confirmed by Xray photoemission spectroscopy [52]. The surface cleanness was confirmed by atomic-force vs. distance curves of SPM, which detected a weak covalent bonding between a Si cantilever and $BaTiO_3$ surface. This showed that atoms of the surface interacted directly with atoms of the SPM tip without intermediate atoms, meaning cleanness in an atomic length scale.

**Prevention of O-vacancy formation during measurements.** As for reduction, UHV is same as ordinary vacuum; No difference in reduction was found between UHV ($< 10^{-7}$ Pa) and high vacuum ($10^{-4} \sim 10^{-3}$ Pa), because the reduction of elements such as oxygen from surface is in non-equilibrium and limited by the evaporation speed of the elements. Heating to 470 K in vacuum is commonly used as a cleaning for Transmission microscopy TEM and considered not to change the $BaTiO_3$ and other oxides. We found that the heating to 470 K increased $\sigma_{T-T}$, indicating that O-vacancy formed by 470 K and $\sigma_{T-T}$ was more sensitive to O-vacancy than TEM. In the present experiments, the highest temperature in vacuum was 410 K for TSSG crystals and 390 K for flux crystals.

The main text showed that BaO-terminated surfaces changed to $TiO_2$-terminated surfaces by heating (~410 K) in UHV, consistently with the literature [40-42]. Contrarily, $\sigma_{T-T}$ of $TiO_2$-terminated surfaces did not change much even by 10 times of heating (~410 K) in UHV (Fig. S9), although repeated electrical mechanical stress of poling, when repeated > 20 times induced visible changes.

**Electrical measurements.** Each of two UHV systems consisted of a treat and a main chamber. In UHV-SPM (JEOL 4610, customized), base pressure was $2 \times 10^{-7}$ Pa (treat) and $2 \times 10^{-8}$ Pa (main). Conductance, capacitance, and nonlinear capacitance were measured through a switch (Keysight E5250A, E5252A) with DC source monitor unit (ADC R6245), LCR meter (Keysight 4284A), and vector signal analyzer (Keysight 89441A), respectively. The laser of SPM was powered off in conduction measurements, although conductance was the same for on and off of the laser.

Conductance $\sigma$ from law $IV$ hysteresis loop curves such as Figs 1(h) were obtained as follows. Current $I$ of ferroelectric consists of current due to the transport of carriers by applied voltage (V) $I^V$, pyroelectricity $I^P$, dielectric current due to dielectricity and dielectric response of space charge limited conduction $I^D$, domain motion $I^{DM}$, current increase by resistance switching $I^R$, and other carrier emission $I^E$. The average $I^{ave}$ (= $(I^+(V) + I^-(V))/2$) vs. $V$ can remove $I^D$, where $I^+(V)$ and $I^-(V)$ are $I$ at a given $V$ during increasing and decreasing $V$, respectively. The automatic fitting to $I^{ave}$ vs. $V$ represents $I^V + (I^P + I^R + I^{DM} + I^E)$, where $I^V$ and $I^P + I^R + I^M + I^E$ correspond to conductance $\sigma$ and shift current, respectively. Additionally, each $IV$ hysteresis loop curve was manually examined. When $IV$ hysteresis was deformed by shift current, the automatically obtained $\sigma$ was deleted or replaced by $\sigma$ reestimated by appropriate fittings, e.g. a fitting to a part of data unaffected by the shift current. The dielectric current $I^D$ may be confused with resistance switching. But, the shape of $I^D$ vs. $V$ is different from that of $I^R$ vs. $V$, $|I^+(V)| > |I^-(V)|$ always for $I^D$, while $|I^+(V)| > |I^-(V)|$ at least one polarity of $V$, and, hence, $I^D$ is distinguishable from $I^R$. $I_{T-T}$'s in Figs. 3(e) and 3(j) are $I^{ave}$'s. In all the present data, no contribution of the resistance switching was found. The absence of the contribution of charge-up to $\sigma_{T-T}$ was confirmed by measuring $\sigma_{T-T}$ before and after shorting all electrodes.

**Scanning probe microscopy SPM.** The accuracy of Kelvin force microscopy (KFM) (JEOL 4610 UHV-SPM) for potential measurements were 5% for −9.5V~ 9.5V. Fluctuation of 20 mV was superposed, which was due to fluctuation or variation of surface charges. PFM was operated in off-piezoelectric resonance in contact mode, and the reliability of PFM was confirmed by piezoelectric hysteresis loops tests. As a test of cleanness of the UHV chamber, clear images of individual Si atoms of a Si [111] surface by non-contact AFM were unchanged for a month, showing the retention of cleanness for that period. $\sigma_{T-T}^{unpoled}$ of BaTiO$_3$ kept in the UHV chamber at 300K was unchanged at least for a year. Vacuum gauge was powered off during all measurements including SPM.

**Properties at cryogenic $T$.** BaTiO$_3$ undergoes two successive 1$^{st}$ order phase transitions below 300K; the lattice symmetry is tetragonal at 300 K but changes to orthogonal and then to rhombohedral with decreasing $T$. Because of these phase transitions, the alignment of $P_S$ was reduced, which substantially decreased $\sigma_{surface}$. The flux single crystal is less pure than TSSG, which may have also decreased $\sigma_{surface}$ by trapping by impurities. Nonetheless, $\sigma_{surface}$ of $P-$ and, especially, $P+$ surface was always much higher than $\sigma_{T-T}$ of $a/c$ surface that was due to bulk conduction, which indicated the persistence of $e^-$ and $h^-$ layer in low-$T$ phases down to 20 K.

**DFT and hybrid functional.** (See also SM-8, SM-9) To date, the novel polar orders that are found here have not been reported, because the DFT of ferroelectric ATiO$_3$ in vacuum with full relaxation of ion position, i.e. geometry optimization requires an unusually large cell and has not reported (A: alkali earth element). To the best of our knowledge, the studies on ATiO$_3$ surface are on those of paraelectric phases except for the one that relaxed only the topmost 1 unitcell and used the unitcell of bulk ferroelectric BaTiO$_3$ phase for the rest of 3 unitcells [53]. On the other hand, we used 41.5 unitcells for BaTiO$_3$/vacuum and BaTiO$_3$ with 80 unitcells for $P+P+/P-P-$ DBs to obtain ferroelectric phase after relaxation. $P_S$ at the location of $e^-h^+$ was almost the same as $P_S$ of a bulk part (Fig. S13). This means that these layers are ferroelectric metals or polar metals.

DFT and hybrid functional calculations were performed with VASP [54], the projector augmented wave (PAW) method [55], and an energy cutoff of 650 eV. All the calculated forces were < 5 meV/Å after geometry relaxation. Local $P_S$ was obtained by a semi-empirical formula with field-correction [56] that agree excellently with Berry phase calculations. Both PBEsol [57] and PBE [58] with Hubbard $U$ [46,47, 59,60-62] were used as exchange-correlation functions. $BaTiO_3$ with O-vacancy in vacuum was calculated similarly with and without spin polarization. Hybrid functional calculations used HSE functional [63,64]. These DFT calculations used mostly graphic processing units acceleration [65,66]. Atomic positions were drawn by VESTA [67]. All the ion positions were fully relaxed except for the length of the slab of $BaTiO_3$/vacuum. The energy minimum state of single-domained defect-free $BaTiO_3$ in vacuum was ferroelectric with free $e^-$ and $h^+$ layers, when $BaTiO_3$ thicknesses was greater than 10 nm. The main locations of $e^-$ and $h^+$ were $Ti_{3d}$ and $O_{2p}$ of $TiO_2$ planes, respectively (also in BaO and SrO planes for Ba and Sr terminated surfaces).

The $a$-axis constant of tetragonal $BaTiO_3$/vacuum was wider by 0.05% than the theoretical bulk value, whereas $BaTiO_3$ is rhombohedral near 4 K. In Fig. 4, defect-free $BaTiO_3$ with a 9-nm-long vacuum was modelled by $1\times1\times41.5\times BaTiO_3$ ($Ba_{41}Ti_{42}O_{125}$: $TiO_2$-termination, $Ba_{42}Ti_{41}O_{124}$: BaO-termination).

O-vacancy is important for reducing atmospheres, and we examined only $TiO_2$-terminated $BaTiO_3$ using $3\times3\times10.5\times BaTiO_3 - O$ ($Ba_{90}Ti_{99}O_{287}$, vacuum length: 5nm). The free energy of $BaTiO_3$ in vacuum is lower with an O-vacancy at $P-$ surface (Fig. 4) than with an O-vacancy at $P+$ surface. The free energy of $1\times1\times80\times BaTiO_3$ without the DBs was always lower than $1\times1\times80\times BaTiO_3$ with the DBs. $BaTiO_3$ with $P+/P+$ and $P-/P-$ DB without and with defect was modelled by $1\times1\times80\times BaTiO_3$ ($Ba_{80}Ti_{80}O_{240}$) and $1\times1\times80\times BaTiO_3 - O$ ($Ba_{80}Ti_{80}O_{239}$).

All the DFT calculations of defect-free $BaTiO_3$ with thickness > 10 nm in vacuum yielded the properties of similar to Figs. 4 and S13. Therefore, the results of Fig. 4 should be valid also in the macroscopic limit at least qualitatively; for example, $\Delta\phi$, buckling, and DOS near the surfaces are almost unchanged in the macroscopic limit. Intrinsic $P+/P+$ DB and $P-/P-$ DB at free surface were found only for large unitcells, which would be the reason why this has not been reported.

In the cases of $SrTiO_3$ in vacuum modelled by $1\times1\times31.5\times SrTiO_3$ ($Sr_{31}Ti_{32}O_{95}$ and $Sr_{32}Ti_{31}O_{94}$), all the ion positions were relaxed except for the length of the slab and the $a$-axis constant that was fixed at 95 % of a theoretical bulk value to stabilize ferroelectricity [68] (Fig. S14).

Three-dimensional (3D) models including the $P+$ and $P-$ domains (e.g., Fig. 3(a)) can disclose defect-free BaTiO$_3$ in vacuum at true free-energy minimum, while the inclusion of a/c domains may be needed. However, because surfaces of such a BaTiO$_3$ consist of $P+$ and $P-$ surfaces, the present results by one-dimensional (1D) model are considered to catch the essential features of the defect-free BaTiO$_3$ in vacuum at true free-energy minimum. The calculation of defect-free BaTiO$_3$ with $P+/P+$ and $P-/P-$ at true free-energy minimum requires 3D models to include parallel and *a/c* domains (Fig. 3(a)). Nonetheless, the state without $P+/P+$ or $P-/P-$ always corresponds to true free-energy minimum at least at 0 K, when defects are absent.

**Polaronic trapping (DFT, HSE).** Defect-free BaTiO$_3$ in vacuum was calculated with both PBEsol and PBE+$U$ ($U$ = 1.3 eV on Ti$_{3d}$ and $U$ = 5 eV on O$_{2p}$, $J$ = 0), to examine polaronic trapping [46,47, 60-62]. Both exchange-correlation functionals yielded the same results except for the slightly better visibility of polaron-like $h^+$ at $P-$ TiO$_2$-surface by PBE+$U$ (The results in Figs. 4(a)-4(d) were obtained with PBE+$U$). Here, $U$ = 1.3 eV on Ti$_{3d}$ (for $e^-$) is much smaller than the standard value of effective $U$ (= $U - J$) on Ti$_{3d}$ [47,60,61]. This is because the previous studies are only on non-ferroelectric Ti-oxides and we find that the standard value of $U$ destroys ferroelectricity. $U$ = 5 eV on O$_{2p}$ (for $h^+$) is slightly smaller than the standard value of effective $U$ (6~8 eV) on O$_{2p}$ [46,62]. This is because the previous studies are only on non-ferroelectric Ti-oxides and we find that the theoretical properties deviate from the experimental ones by the standard values of $U$. Therefore, the present results may not capture the properties of polarons quantitatively. However, the comparison with these results with the results with PBEsol can show whether the polaronic effect changes conduction properties, at least in case of $h^+$; The results show free $e^-h^+$ instead of trapped $e^-h^+$.

Further, TiO$_2$-terminated BaTiO$_3$ in vacuum was calculated with HSE that can correctly estimate polarons by fixing the middle part of a BaTiO$_3$ slab at the geometry of bulk BaTiO$_3$ (for computational costs). The results were similar to the PBEsol and PBE+$U$ calculations and showed free $e^-h^+$ instead of trapped $e^-h^+$.

**Surface structures** [42,43,53,69-73]**.** Outward polarization or outward dipole at surfaces of the paraelectric (cubic) ATiO$_3$ has been extensively studied for ideal surfaces with one-dimensional (1D) geometry [42,43,53,69-73], similar to the geometry of our DFT. On the other hand, surface structures such as (2×1) of TiO$_2$ surface of heavily reduced semiconducting BaTiO$_3$ are studied by DFT with 2D geometry [53,72,73]. Opposite to the samples used in conventional studies [72,73], our BaTiO$_3$ surface was cleaned by heavy oxidation that enabled over-oxidation. Chen et al. [53] predicted that the (2×1) surface was metallic even without additional defects, whereas the unpoled surface of our BaTiO$_3$ was not ($\sigma_{surface}$ = 0). Therefore, our DFT did not include such surface structures like (2×1) and followed the conventional approaches [42,43,68-70].

We found only one literature that studied surface polarization or dipoles of ferroelectric phase BaTiO$_3$ using 4 unitcell height models [53]. For such a short BaTiO$_3$, DFT should yield paraelectric phase when all the ion positions are relaxed [38]. Similarly, the ion positions of defect-free BaTiO$_3$ with $P+/P+$ and $P-/P-$ DB was fixed at those of bulk ferroelectric phase BaTiO$_3$[38].

**SM-3 Effective thickness of inplane bulk conduction**

The conduction paths in the absence of $e^-$ or $h^+$ surface layer, i.e. the electric field lines are drawn in Fig. 1(e). Because the field lines of DC field are approximately the same for those of capacitance measurements, the effective thickness was estimated from the capacitances of the inplane (T1-T2) and bulk (T1-B, T2-B) for each electrode configuration. This value was determined by the gap width and the width of electrodes. The effective thickness was close to the doubled value of the distance between the centers of T1 and T2: 60 μm for Fig. 1 and Fig. 3 for 40 μm, which were about 1/10 of the nominal thickness of the samples.

**SM-3 Additional evidence of unimportance of defects in Figs. 1-3**

For the efect density $N_t$ (4×10$^{17}$/cm$^3$) in SM-1 Samples, the surface charge density due to defects is 0.6×10$^{-6}$ Ccm$^{-2}$ and only 3 % of $P_S$, when all the defects and impurities in a 100 nm thick layer are assumed to be singly charged. Therefore, the contribution of $N_t$ to screening is negligible. Actually, $\sigma_{T-T}$ (as well as surface potential) of $a/c$ domains remained low in UHV at 300 K over a year: 5~50 pΩ$^{-1}$ (flux crystals) and 1~10 pΩ$^{-1}$ (TSSG crystals). Further, $\sigma_{T-T}$ of $P+$ surface of Fig. 1 and Fig. S9(c) decreased after kept in UHV for 10 h and also for 12 days at 300 K [8], whereas oxygen reduction should increase $\sigma_{T-T}$. Moreover, $\sigma_{T-T}$ of both $e^-$ and $h^+$ of TiO$_2$-terminated surface was unchanged by the repletion of heating-cooling without and with poling in UHV (Fig. S9).

**SM-5 Effect of anisotropy of mobility**

An alternate explanation of the changes of $\sigma_{T-T}$ (Figs. 1-3) may be the anisotropy of the mobility. The ratio of the mobility along $a$-axis to that along $c$-axis is 9.2 for $e^-$ and 19.6 for $h^+$ [74]. First, the ratio of $\sigma_{surf}$ of $P+$ surface ($e^-$) to $a/c$ state was much larger than this anisotropy. Second, the $T$- dependence of $\sigma_{surf}$ of $P+$ and $P+$ surface was different from that of the anisotropy [74]. Third, inplane capacitance sometimes indicated that $a$-axis of $a/c$ surface was aligned along the inplane conduction path, but the inplane conduction of the $a/c$ state was much lower than that of $P+$ and $P+$ state (near 300 K). Therefore, the anisotropy was not the origin of the observed changes of $\sigma_{T\text{-}T}$.

**SM-6 Potential by standard theory**

Standard theories [9] consider only dielectric screening without $e^-$ and $h^+$, which is the reconfiguration of polarization distribution. In Fig. S11, potential and polarization are calculated with grounded electrodes by the same formalism as Kittel model [9] that produced nano-domains and vortex-like [10, 14] $P_S$ distributions [75]. Both analytical approach (Fig. S11(a)) and finite element simulation (FEM) (Fig. S11(b)) showed the similar results. Initial $P_S$, thickness, width of free surface (gap), and relative permittivities $\varepsilon_a$ and $\varepsilon_c$ were 20 $\mu$C/cm$^2$, 220$\mu$m, 5$\mu$m, 1000 and 200 for FEM. The parameters except for width (30 $\mu$m corresponding to Fig. 3) were used for analytical results. The magnitude of the potential was almost unchanged for typical ranges of experimental values of $\varepsilon_a$ and $\varepsilon_c$.

Kittel-like models minimize free energy by optimizing only polarization distribution and neglecting associated mechanical strain [14]. Therefore, the electrostatic energy minimization is not restricted by strain, while those of DFT are [10]. Therefore, Kittel-like models approximately yield polarization configurations that correspond to the lower bound of electrostatic energy. This implies also that the electric field and the potential given by Kittel-like models are approximately the lower bound. However, the magnitude of potential in Fig. S11 far larger than experimental ones (Fig. 3), which could not be resolved within the framework of Kittel-like models (we tested various parameters). Additionally, the standard theories including Kittel models cannot explain the stable existence of wide $P+$ or $P-$ single domains (30 $\mu$m × 1000 $\mu$m) in Figs. 1-3 [9, 14,75]. On the other hand, this is explicable by $e^-h^+$ layers, of which formation energy is a tiny fraction of electrostatic energy given by these models [8,17].

**SM-7 Depletion layer and *pn* junction at *P*+/*P*− boundary**

When the screening by defects are negligible, $P_S/e$ divided by an effective thickness $l_e$ is regarded as an effective ionized dopant density per area [76]. Therefore, *pn* junctions form at the boundaries of $P+$ and $P-$ surface [71] as in Fig. 3(a), which is the origin of the low conductance of unpoled state in Fig. 3(e). Similarly, with increasing $T$, the alignment of polarization decreases. This means that $P+$ surface starts to contain many $P-$ domains and $pn$ junctions and explains the rapid decrease of conductance around $T^*$ in Fig. 1(j).

In the depletion layer of a 1D Schottky contact, potential $\phi$ is $eNx^2/2\varepsilon\varepsilon_0$[1], where $N$, $x$, $\varepsilon$, and $\varepsilon_0$ are ionized impurity density, distance from the metal-semiconductor boundary, permittivity, and vacuum permittivity, respectively. When we include charged defects or traps with density $N_{defects}$, the effective ionized dopant density per area $N_{eff}$ is $P_S/l_e e \pm N_{defects}$, where defects or traps are assumed to be singly charged. Therefore, $\phi = eN_{eff} x^2/2\varepsilon\varepsilon_0$ in a depletion layer, and $\phi$ is constant outside. Here, $N_{eff} < |P_S|/l_e e$, because the charging of trap or the migration of charged defects are considered to screen $P_S$.

This formula was used in the fitting to $\phi$ of the $P+$ surface in Fig. 3(d) and $\phi$ of the $P-$ surface in Fig. 3(i), showing that abrupt change in $\phi$ was due to depletion layers. This fitting yielded $N_{eff} \approx 0.2 \sim 0.6 \times 10^{16} \varepsilon$ (cm$^{-3}$), where $\varepsilon = 500 \sim 2000$ in the direction perpendicular to $c$-axis. The obtained $N_{eff}$ is $10^{18} \sim 10^{19}$ cm$^{-3}$, which corresponded nearly to a degenerate semiconductor. Therefore, the effect of Schottky contact was not considered as severe, as seen in the ohmic conductance of the $P+$ surface in Fig. 3(e) and the $P-$ surface in Fig. 3(j).

**SM-8 Experimental vs. theoretical potential**

The abrupt change of the potential at the electrode-BaTiO$_3$ boundary (Fig. 3) is due to depletion layer formed by a contact potential $-e\phi_C = E_{BTO} - E_M$, where $E_{BTO}$ and $E_M$ are Fermi levels $E_F$'s of BaTiO$_3$ and the electrode measured from the vacuum level, respectively. As shown below, the theoretical difference in $\phi_C$ between TiO$_2$ and BaO termination is $\Delta\phi_C^{cal} = \phi_{TiO}^{cal} - \phi_{VTiO}^{cal} - (\phi_{BaO}^{cal} - \phi_{VBaO}^{cal})$, where $\phi_{TiO}^{cal}$, $\phi_{BaO}^{cal}$, $\phi_{VTiO}^{cal}$, and $\phi_{VBaO}^{cal}$ are the theoretical potential of BaTiO$_3$ and vacuum for TiO$_2$ and BaO termination in Fig. 4, respectively, Here, the values of $\phi_{TiO}^{cal}$ etc. are the values of the midpoint of BaTiO$_3$ and vacuum of Figs. 4(a) and 4(c), respectively, representing the average value of each part.

Using the above expression of $-e\phi_C$, the difference $\Delta\phi_C$ is $-e\Delta\phi_C = E_{TiO} - E_M - (E_{BaO} - E_M) = E_{TiO} - E_{BaO}$, where $E_{TiO}$ and $E_{BaO}$ are $E_F$'s of TiO$_2$- and BaO-terminated BaTiO$_3$ measured from the vacuum level, respectively. Here, we have assumed that the potential or $E_M^{exp}$ remains unchanged by poling, because the electrode is grounded. In the bulk limit, *i.e.* a thick BaTiO$_3$, theoretical $E_F$'s relative to vacuum level $E_{TiO}$ and $E_{BaO}$ are equal to $-e(\phi_{TiO}^{cal} - \phi_{VTiO}^{cal} + \delta)$ and $-e(\phi_{BaO}^{cal} - \phi_{VBaO}^{cal} + \delta)$, respectively, where $\delta$ is the offset between $E_F$ and the highest potential ($\sim$ potential of the outermost orbit) at the midpoint of BaTiO$_3$. By substituting these equalities into $-e\Delta\phi_C = E_{TiO} - E_{BaO}$, the above expression of $\Delta\phi_C^{cal}$ is obtained. This means that $\Delta\phi_C$ between TiO$_2$ and BaO termination ($E_{TiO} - E_M - (E_{BaO} - E_M)$, Fig. 3) corresponds to $\Delta\phi_C^{cal}$ between TiO$_2$ and BaO termination (Fig. 4).

### SM-9 Potential difference vs. bandgap (Fig. 1a): Hybrid functional

Figure 1(a) assumed that the difference in potential $e\phi$ ($e\Delta\phi$) between P+ and P+ surface of defect-free BaTiO$_3$ and SrTiO$_3$ in vacuum was close to $E$g. The validity of this assumption $e\Delta\phi \sim E$g is as follows. In the absence of $E_d$, the top of the valence band at surface $E_V = E_C - E$g ($E_C$: bottom of the valence band at surface) is at $E_F$. By $E_d$, the band bends as indicated by Figs. 3(d) and 3(i) but does not cross $E_F$ much. This is because the band crossing energy $\Delta E$ required for $P_S \sim 2\times10^{-5}$ C/cm$^2$ is approximately 0.1 eV[1]. The conditions of $e^-$ and $h^+$ layer formation are $E_C^+ = E_F - \Delta E$ ($E_C^+$: $E_C$ at P+ surface) and $E_V^- = E_F + \Delta E$ ($E_V^-$: $E_V$ at P- surface), which yields Fig. 1(a). Because $-e\Delta\phi = E_C^+ - E_C^-$ ($E_C^-$: $E_C$ at P- surface) and $E_V^- = E_C^- - E$g, $e\Delta\phi = E$g $+ 2\Delta E \sim E$g. The experimentally deduced band bending at P− and P+ surface well correspond to the band near P− and P+ surface of Fig. 1(a).

In the DFT calculations of Fig. 4, the difference in potential $e\phi$ between P+ and P- surface was 1.5 eV for TiO$_2$ and 1.8 eV for BaO. To correct the error in these estimations of $E$g associated with DFT, bulk [78] and a- and c-face BaTiO$_3$ with BaO and TiO$_2$ termination in vacuum were calculated with HSE [62,63] and PBEsol [59], and PBE [60] with Hubbard $U$ [61]. Here, $E$g of BaTiO$_3$ in vacuum or with surface was always smaller than $E$g of bulk BaTiO$_3$, because of unpaired electron states at surfaces. The bandgap $E$g of bulk BaTiO$_3$ by HSE agreed excellently with experimental $E$g [78], and, hence, HSE was regarded to yield accurate estimations of $E$g of BaTiO$_3$/vacuum. $E$g of bulk BaTiO$_3$ by PBE+$U$ was approximately 1 eV lower than that of HSE. $E$g of a-face BaTiO$_3$ with TiO$_2$ and BaO termination in vacuum by HSE was approximately 0.7 eV higher than $E$g by PBE+$U$. Therefore, in the relative comparison with $E$g, the difference in $e\phi$, 1.5 V and 1.8 V in Fig. 4 corresponds to 2.2 eV and 2.5 eV of HSE. Consequently, the experimental $e\phi$ difference after heating (TiO$_2$ termination) in Fig. 3 quantitatively agreed with the theoretical $E$g of TiO$_2$-terminated BaTiO$_3$. The imperfect agreement in the case of BaO termination was probably due to the incomplete coverage of the surface by BaO terminations converted from the original TiO$_2$ terminations.

### SM-10  Valid time-scale of $e^-h^+$ for ferroelectric properties and $E_d$

We studied ferroelectrics with minimum defects to show the properties of ideal ferroelectrics, as the most theories [6,8,10,14,18,19] consider defect-free ferroelectrics for $E_d$-related phenomena.

However, ferroelectrics in experiments contain defects to degree to change some properties [1-3,8,17-27,32,34,35-37,44,45,48,49,76].

The valid range of the time scale for which the $e^-h^+$ layer is essential for ferroelectric properties depend on the speed of $e^-h^+$ transport, adsorbates, defects, and ion transports. The $e^-h^+$ layers are expected to form by the transports of $e^-h^+$; for example, $e^-h^+$ are supplied from the electrodes in the poling. Therefore, the distance from electrodes and temperature affect the lower bound of the time scale, which separate experiments suggests <10 ms (e.g., Fig. 3(f)). Below this lower bound, the screening of $E_d$ by $e^-h^+$ layers is negligible. On other hand, our and many DFT results neglected the formation and transport of defects by depolarization field $E_d$. Even when the screening by these defects are negligible in short time scale, they should be important in a long time scale. Actually, $\sigma_{T-T}$ of a $P+$ state became 30% of the original value after kept in UHV at 300 K for 12 days [8], which was attributed to adsorbates, the decrease of alignment of $P_S$ by $E_d$, and migration of defects. Consequently, the present results are considered valid in a middle time-scale. The validity up to ~10h was in ultraclean environments (Figs. 2(b) and S10). The upper bound time scale for the screening of $E_d$ by $e^-h^+$ decreases drastically with uncleanness of environments and the increase of the defects in samples. Therefore, $E_d$ is considered as practically absent in experiments including those attributing the results to $E_d$ [11-13].

## References (Supplemental Materials)

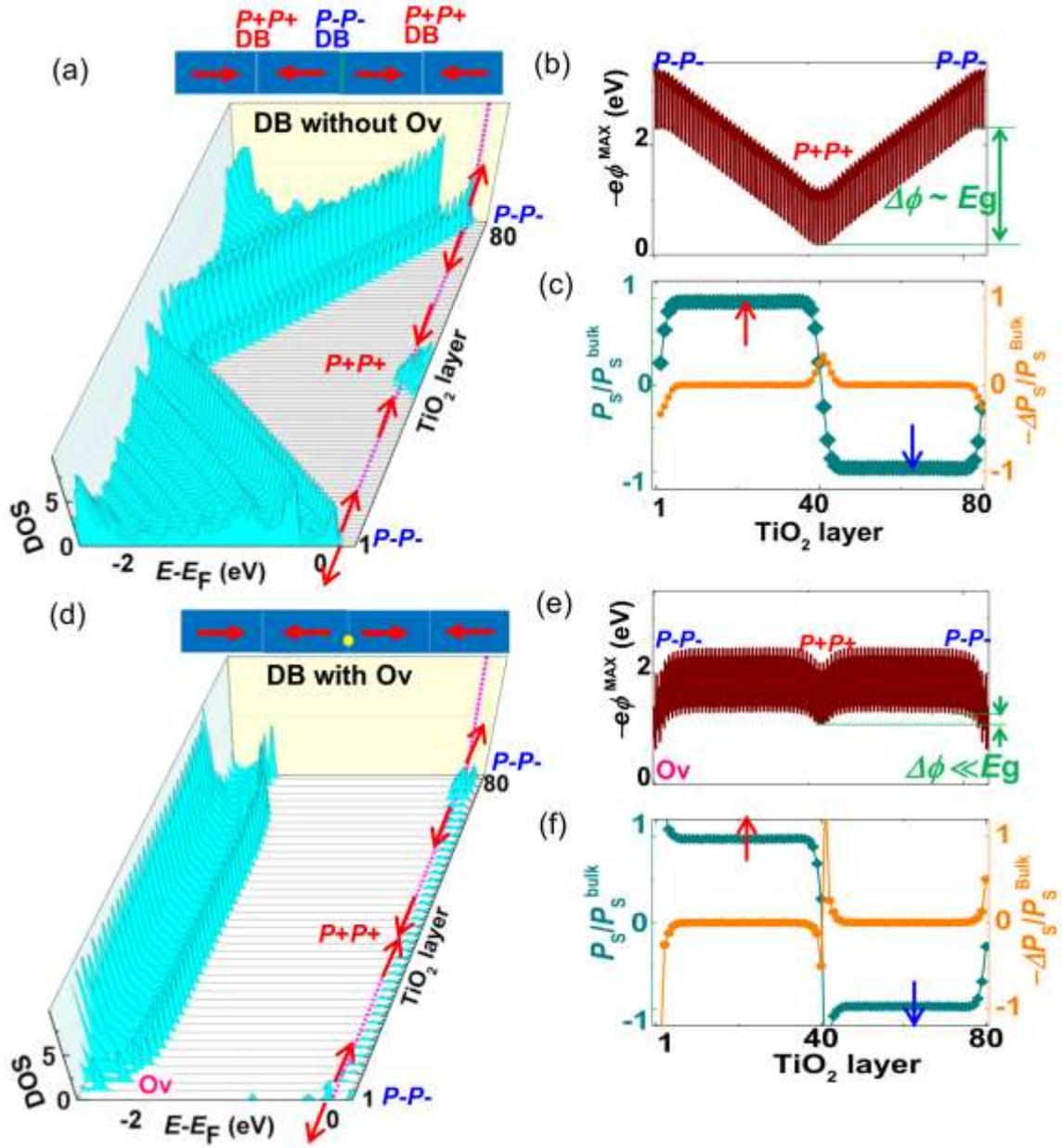

FIG. S1 BaTiO$_3$ having a $P+P+$ and a $P-P-$ DB by DFT. (a)-(c) Without oxygen-vacancy (Ov). (d)-(f) $P_S$-defect complex with Ov in BaO layer at $P-P-$ (0.4% of all O's, Ov at $P+P+$ was unstable and not shown). (a)(d) Local density of states (/eV) of each TiO$_2$ layer along $c$-axis // $P_S$. (b) Potential along $c$-axis // $P_S$ shows $e\Delta\phi \approx E$g. (e) $e\Delta\phi \ll E$g. Red and blue arrows show polarization directions. DOS indicates $e^-$ at $P+P+$ and $h^+$ at $P-P-$. (c)(f) Local $P_S$ and polarization charge $-\Delta P$ normalized by bulk $P_S$ along $c$-axis varied symmetrically, yielding an $e^-$–$h^+$ symmetry in the DOS and indicating that the $e^-$ and $h^+$ conductance were similar in magnitude. With O-vacancy, DOS at $P+P+$ was substantially reduced at $E_F$, while $P_S$ did not change much. These characteristics are consistent with previous experiments [18-19, 22-27]

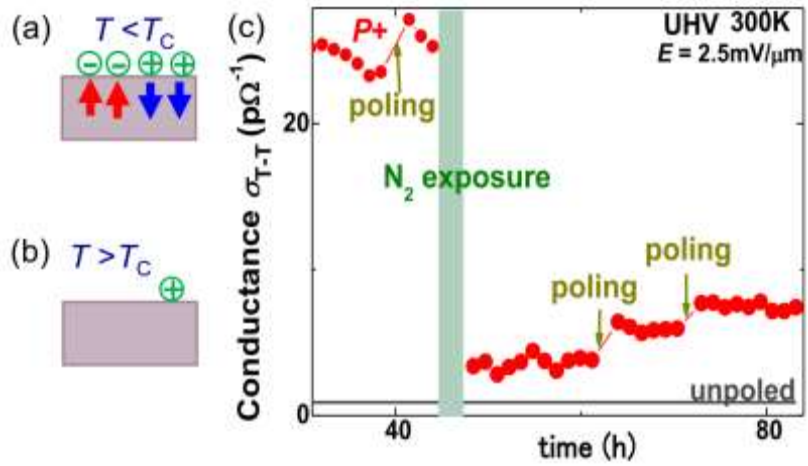

FIG. S2. Charge cleaning (UHV, 300K). (a)(b) Illustration of adsorbate charges at $T < T_C$ and $T > T_C$. (c) Example of field-induced charge cleaning; $\sigma_{T-T}$ (~ $\sigma_{surface}$) after $N_2$ exposure increased stepwise by the instant $P_S$ change (This crystal was KF-flux crystal).

Principle: Charge cleaning is achieved by changing surface $P_S$, because most charged adsorbates fly out by repulsive force from $P_S$ of the polarity opposite to that of the adsorbates. This was done by the electrical switching of $P_S$ or the disappearance of $P_S$ by heating above $T_C$. Here, charged adsorbates are considered to have charges that neutralizes surface $P_S$. Therefore, the breakdown of the charge neutrality by the change of surface $P_S$, e.g. via electrical poling, kicks out the charged adsorbates.

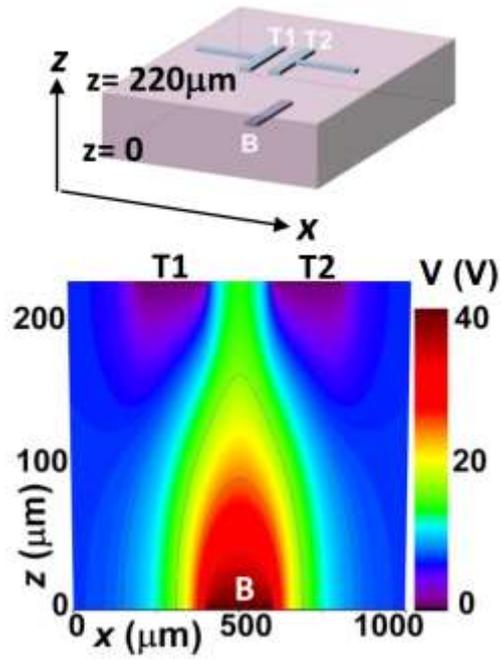

FIG. S3. Potential in xz cross section in paraelectric $BaTiO_3$ during poling (40V), calculated with Kittel model ($P_S$= 0, $\varepsilon_x$ =$\varepsilon_z$ = 3000).

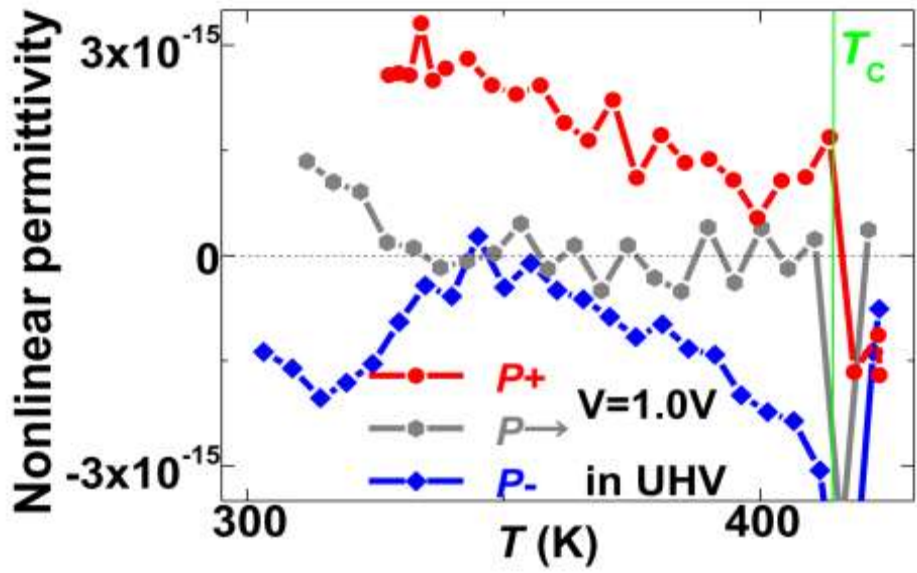

FIG. S4. Nonlinear permittivity $\varepsilon^2$ (F/V$^2$) confirming the orientation of $P_S$ in UHV. Measurements using electrodes T1 and B (This crystal was the same as that in Fig. 1).

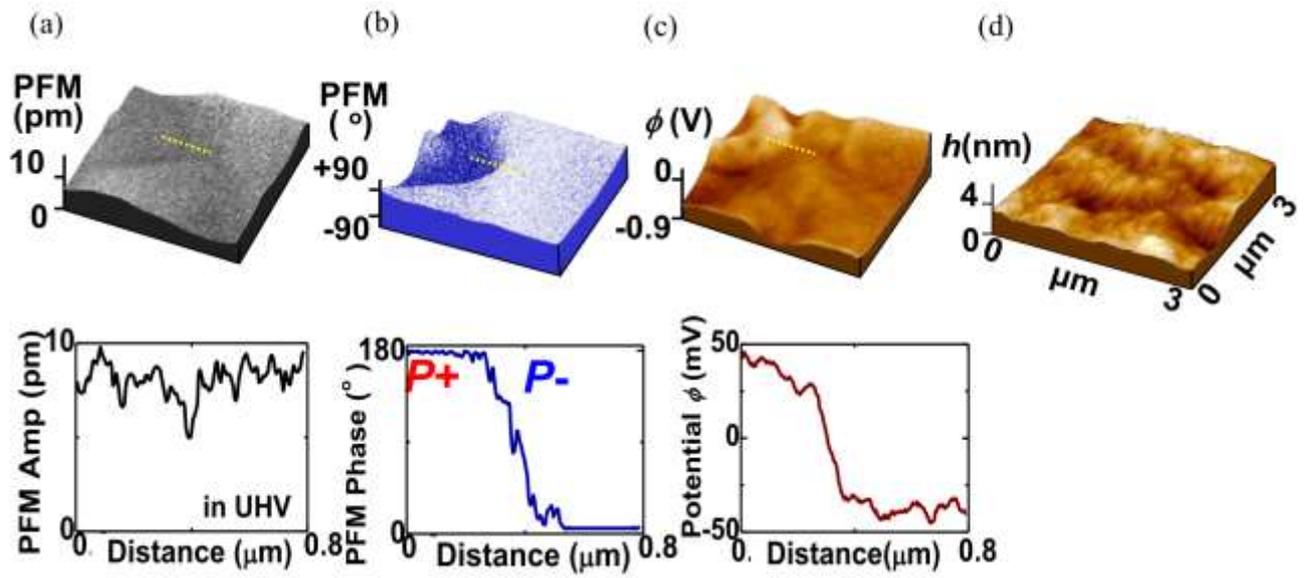

FIG. S5. Correspondence of potential image by KFM to piezoelectric image by PFM of a same location (UHV, 300K). (a) Out-of-plane PFM amplitude. (b) PFM phase. (c) Electrostatic potential. (d) topology. The graph underneath each images is a cross-section analysis along the yellow dotted lines in the image (This crystal was KF-flux crystal).

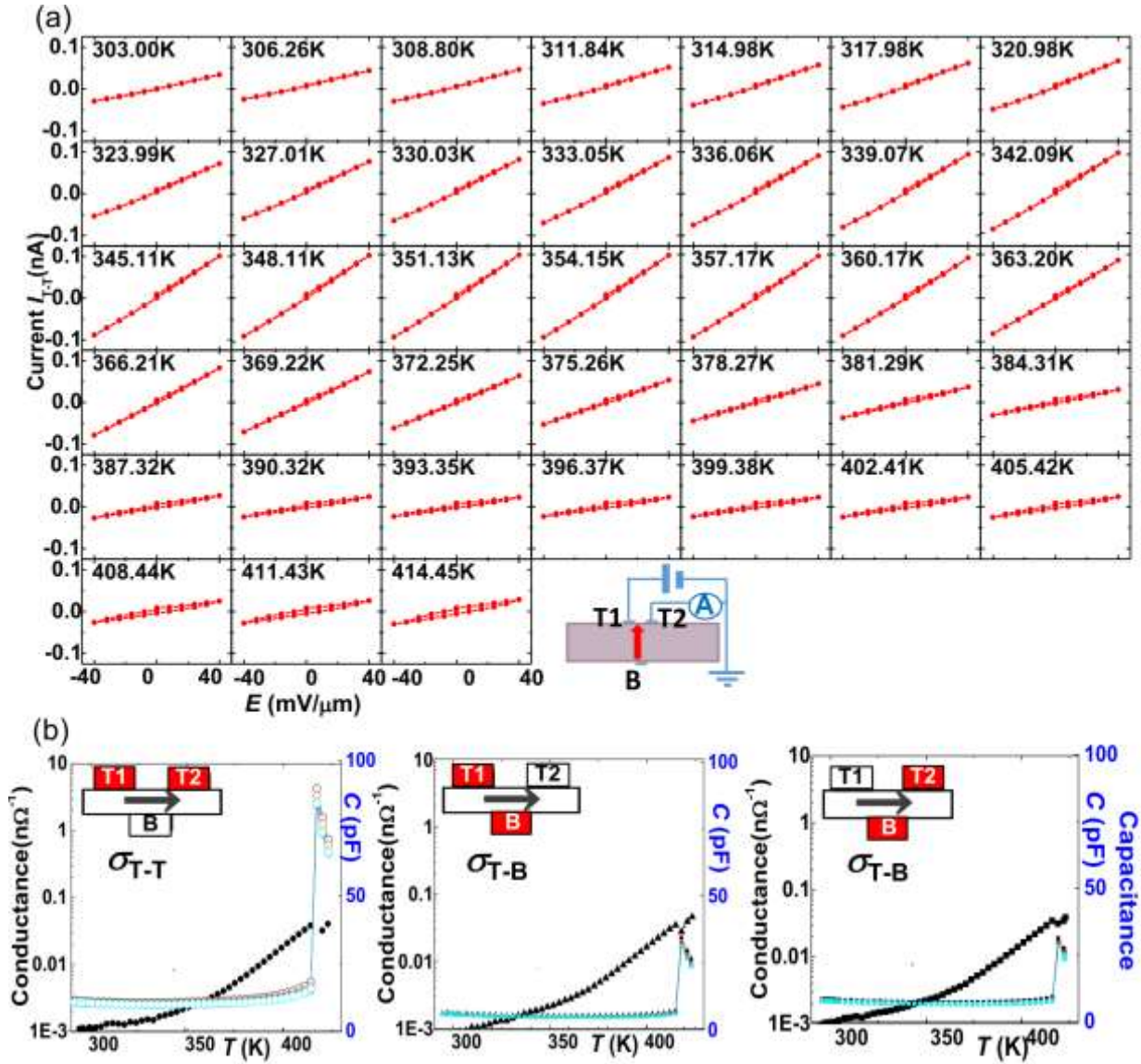

Fig. S6. Conductance measurements: (a) Original *IV* hysteresis loop curves and (b) a standard set of *T*-dependence. (a) Original *IV* hysteresis loops of $\sigma_{T-T}$ of *P*+ surface of Fig. 1(k). (b) Conductance and capacitance of an unpoled state between T1-T2 (inplane), T1-B (bulk), and T2-B (bulk) (measured in all the *T*-dependence measurements).

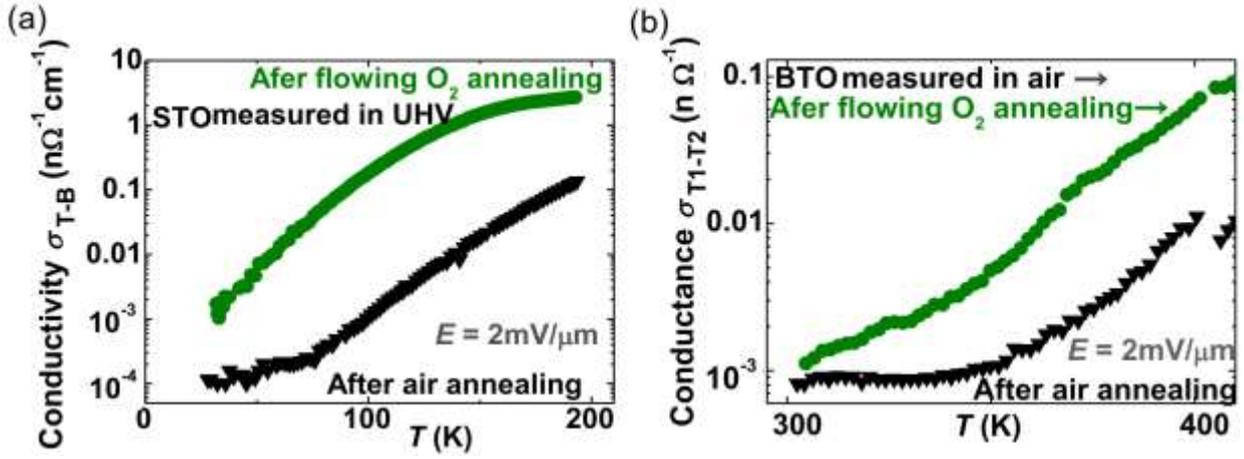

FIG. S7. Bulk conductance of unpoled single crystal after annealing in 1-atm flowing oxygen and air, supporting $h^+$ conduction. (a) SrTiO$_3$. (b) BaTiO$_3$. According to refs. 44 and 45, BaTiO$_3$ and SrTiO$_3$ in annealed in air are considered slightly $h^+$ rich, which is enhanced by oxygenation. Therefore, the results here show that conductance increased by oxygen annealing was due to $h^+$. The conductance after annealing in 1-atm air was the same as those of as-grown or *as-cut* crystals.

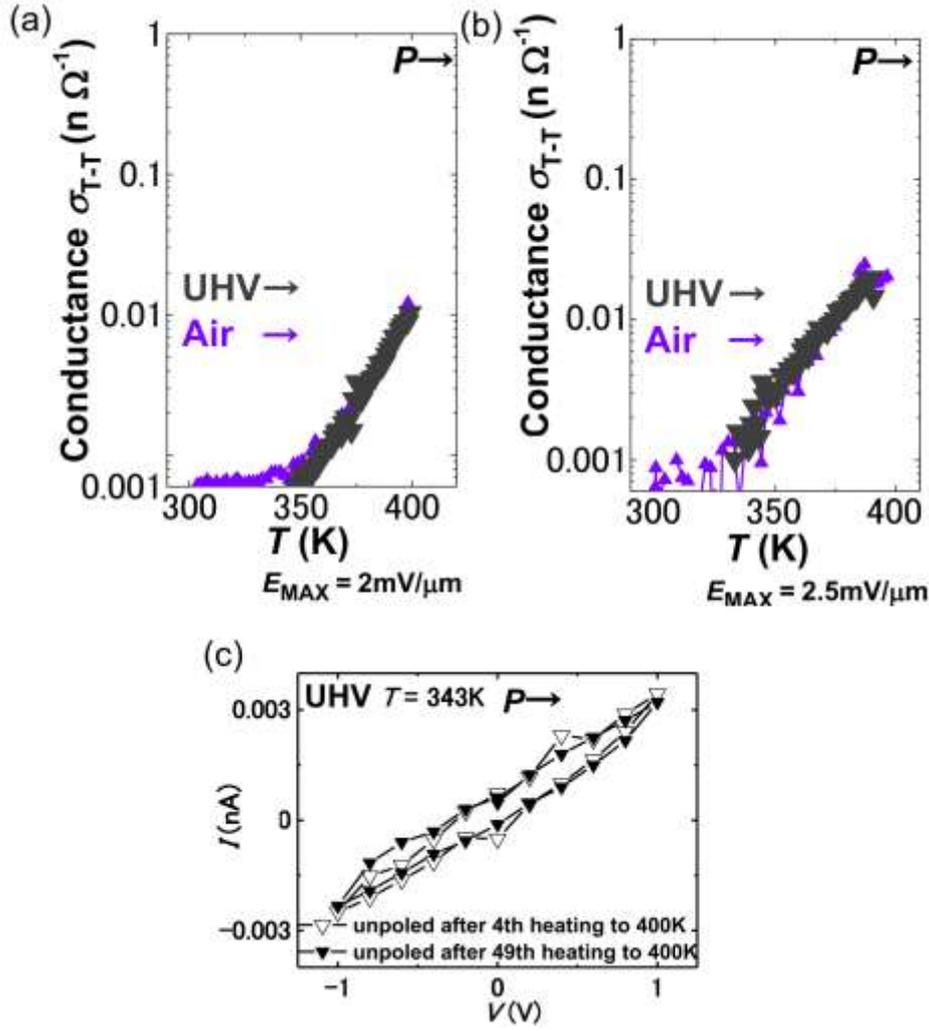

FIG. S8. Comparison of $\sigma_{T-T}$ of two unpoled states measured in air and UHV and insensitivity of $\sigma_{T-T}$ of unpoled state to heating. (a)(b) Conductance between T1 and T2 $\sigma_{T-T}$ of each BaTiO$_3$ was first measured in air and, then in UHV. Both TSSG (a) and KF-flux BaTiO$_3$ crystal (b) show no difference in $\sigma_{T-T}$ between air and UHV.

(c) No difference in $\sigma_{T-T}$ of unpoled state after 4 and 49 times of UHV heating to 400K (KF-flux BaTiO$_3$ crystal). This supports that $\sigma_{T-T}$'s of unpoled states were due to the bulk and the effect of these heatings was insignificant for $\sigma_{T-T}$, because Ov's were expected to form by these heatings.

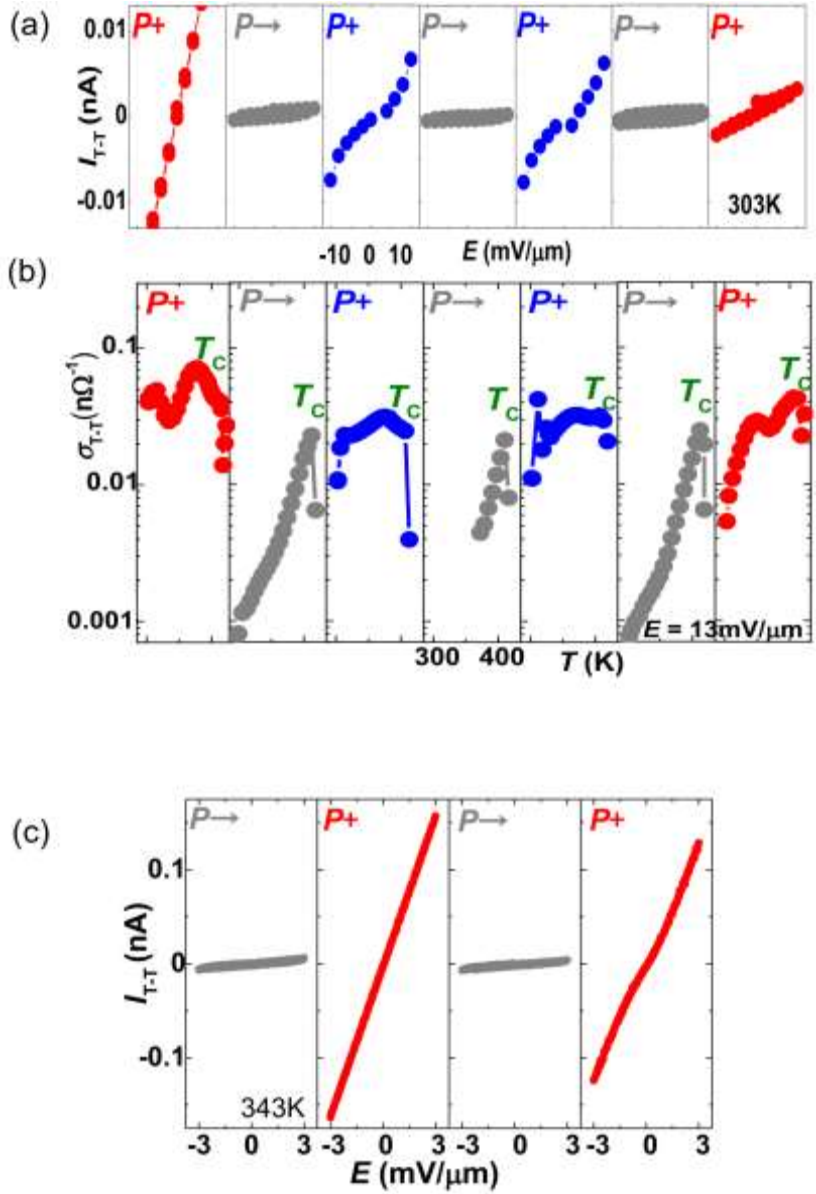

FIG. S9. Repeatability of conductance $s_{T-T}$ showing the dominance of $P_S$ on $\sigma_{surface}$ of BaTiO$_3$ (UHV). BaTiO$_3$ had TiO$_2$-terminated surface layers. All graphs in each panel are in chronological order. (a) $IV$ curves after repeated changes of $P_S$ orientation (average current at each voltage). (b) $T$-dependence of $\sigma_{T-T}$ of the state corresponding to (a). (c) $IV$ hysteresis loops of different poled states of a KF-flux BaTiO$_3$; $\sigma_{T-T}$ of $a/c$ surfaces was unchanged, and $\sigma_{T-T}$ of the $P+$ state was unchanged or slightly decreased in the 2$^{nd}$ ↑.

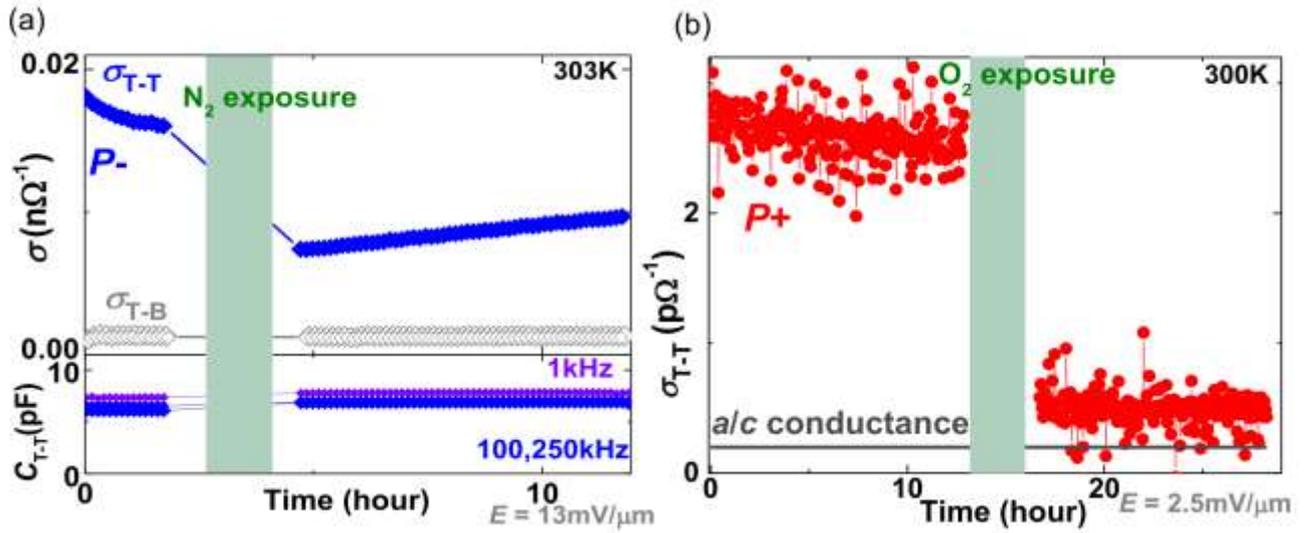

FIG. S10. $\sigma_{surface}$ before and after gas exposure of BaTiO$_3$ having TiO$_2$-outermost surface layers. (a) Conductance and capacitance of a $P-$ state before and after 99.9% N$_2$ gas exposure. The magnitude of the bulk conductance $\sigma_{T-B}$ is always almost the same as $\sigma_{T-T}$ of unpoled state. The capacitance unchanged by the exposure suggests that $P_S$ was unchanged by the exposure. (b) $\sigma_{T-T}$ of a $P+$ surface before and after 99.9% O$_2$ gas exposure. The low conductance is due to the coarse surface of an *as-cut* TSSG BaTiO$_3$ crystal. $\sigma_{T-T}$ of *a/c* surface estimated from separate experiments is shown by grey lines.

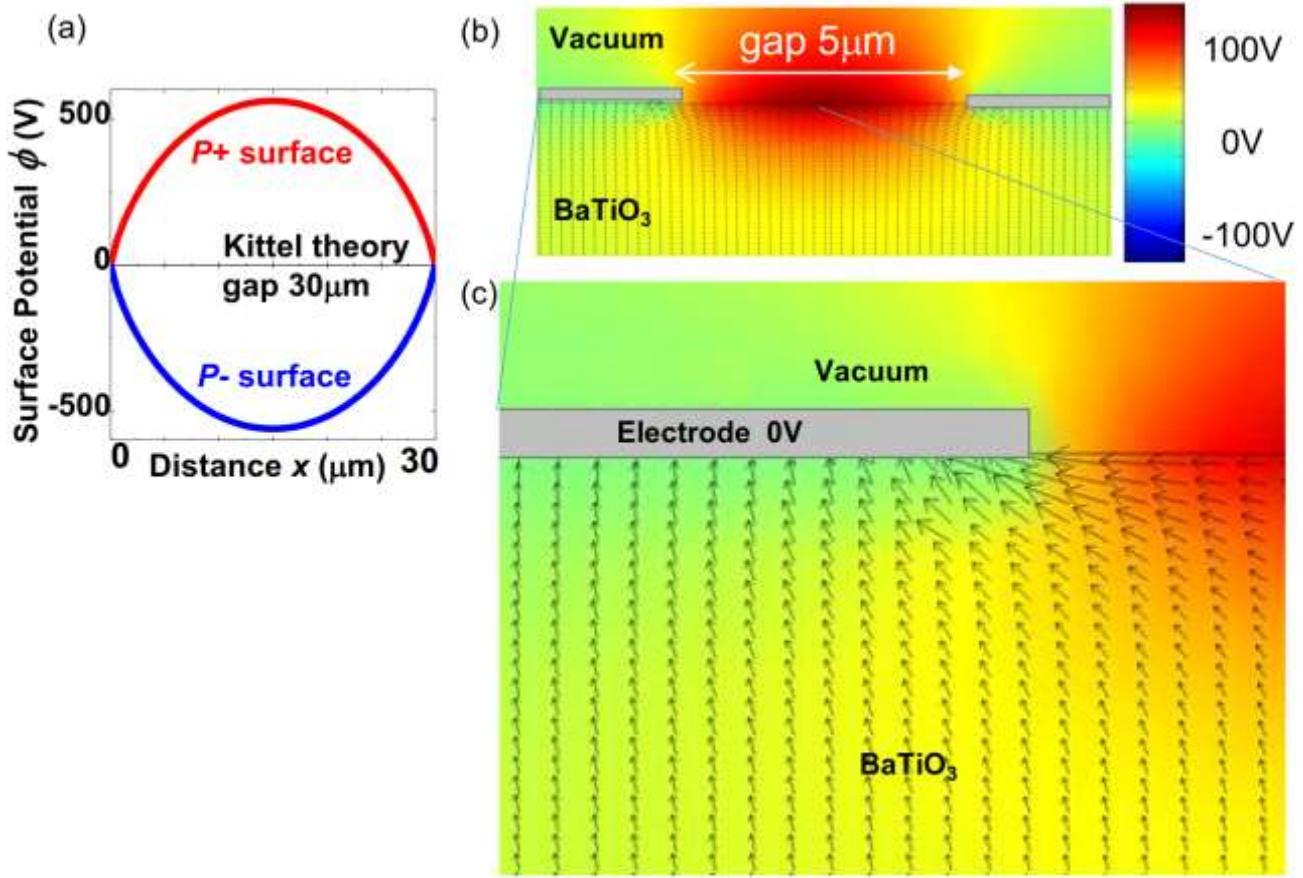

FIG. S11. Potential and $P_S$ of BaTiO$_3$ in vacuum under no external field by Kittel model (See also SM-6). Electrode is grounded. (a) Surface potential by a Kittel model for the electrode configuration of Fig. 3 (Gap between electrodes: 30μm). (b) Potential and $P_S$ distribution (Gap between electrodes: 5μm). (c) Expanded view of (b). $P_S$ orientation and magnitude are shown by arrows.

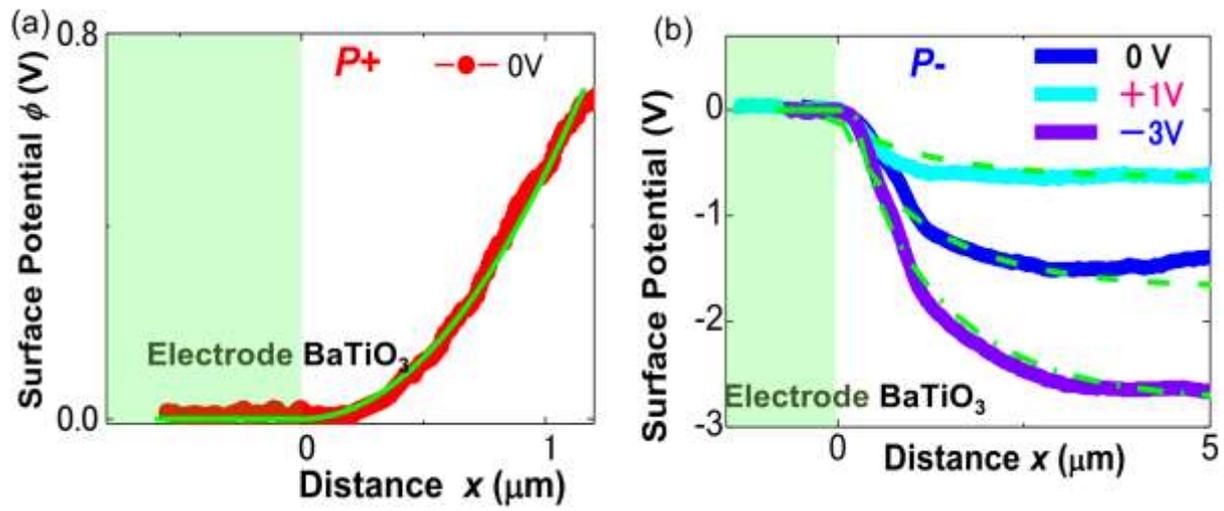

FIG. S12. Surface potential at BaTiO$_3$ surface near electrode and fitting with depletion layer theory (See also SM-7). (a) P+ surface in Fig. 3(b). (b) P- surface after 410K-heatings (Fig. 3(h)) under bias of 0, 1V, and -3V between T1 and T2.

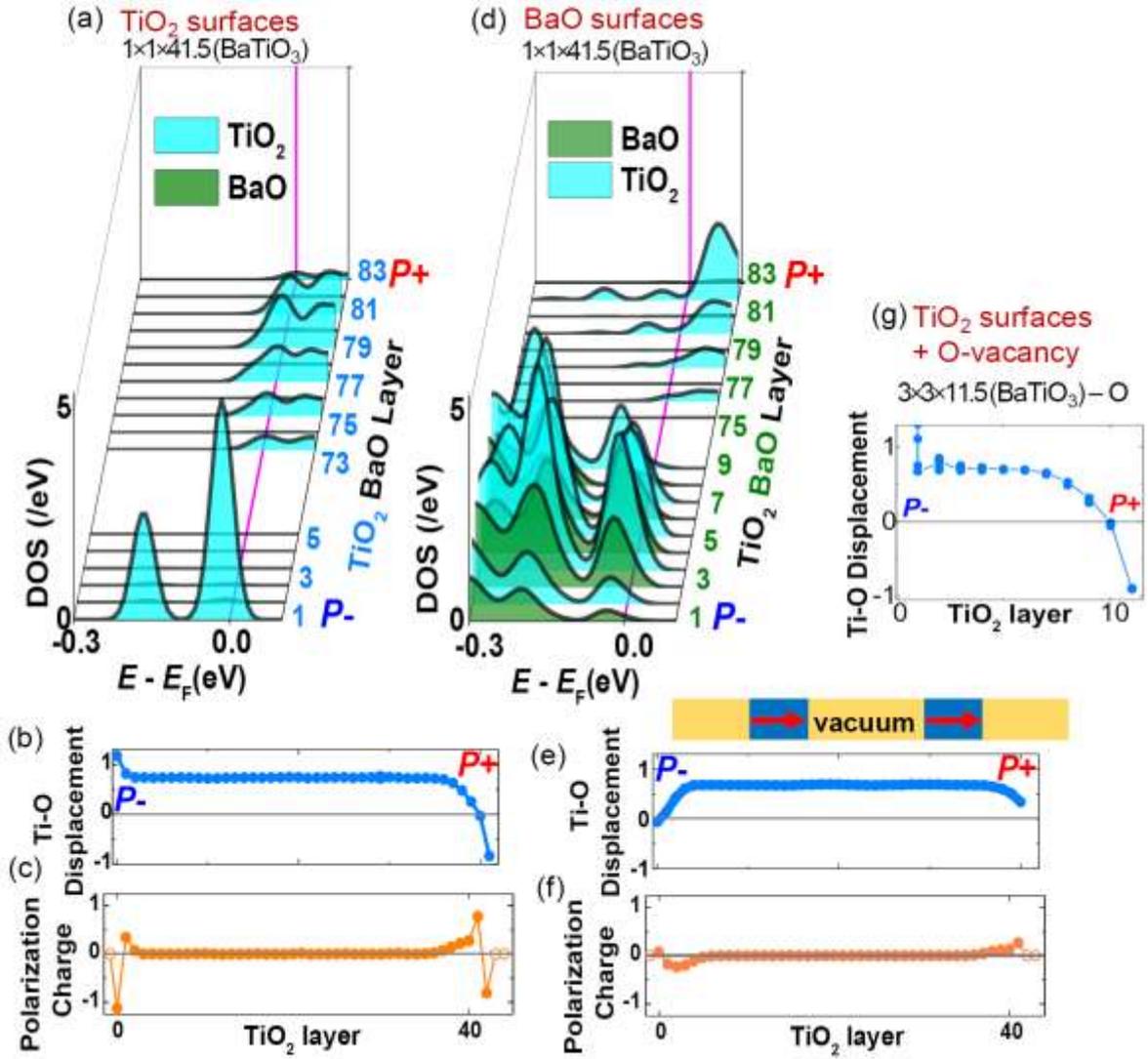

FIG. S13. DOS, buckling at the surface, and $-\Delta P$ of defect-free $BaTiO_3$ having (a)-(c) $TiO_2$- and (d)-(f) BaO-terminated surfaces in vacuum (DFT). (a)(d) Local DOS of $TiO_2$ and BaO layers near top and bottom surface along $c$-axis (// $P_S$). (b)(e) Displacement of Ti from O along $c$-axis in $TiO_2$ plane of each unitcell. (c)(f) polarization charge $-\Delta P$ calculated using (b) and (e). In (b), buckling appears as a change in $\Delta z_{Ti-O}$ from positive to negative near the $P+$ surface ($\Delta z_{Ti-O}$: displacement of Ti from O along the $c$-axis). Because $P_S \propto \Delta z_{Ti-O}$ [32], this $\Delta z_{Ti-O}$ shows an intrinsic $P+P+$ DB, which enhanced polarization charge density, $-\Delta P$, near the $P+$ surface. In (e), $\Delta z_{Ti-O}$ changed from negative to positive at the $P-$ surface but remained positive near the $P+$ surface. This yielded a $P-P-$ DB near the $P-$ surface and a large $-\Delta P$ there, and a small $\Delta P$ near the $P+$ surface). (g) Displacement of Ti from O along $c$-axis in $TiO_2$ plane of each unitcell of $TiO_2$-terminated surface with an O-vacancy at $P-$ surface.

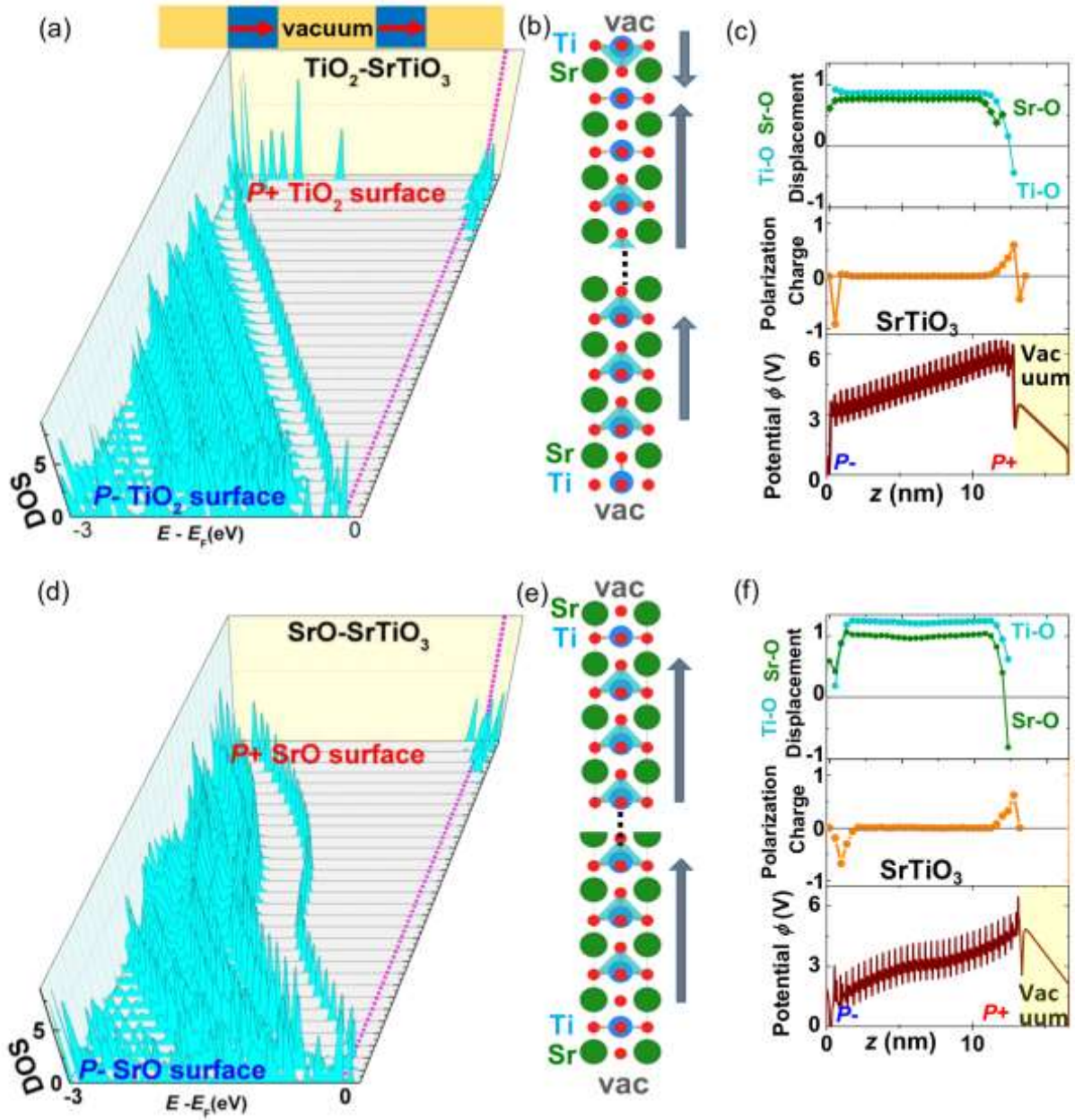

FIG. S14. Strained defect-free SrTiO$_3$ with (a)-(c) TiO$_2$- and (d)-(f) SrO-terminated surface in vacuum (DFT), indicating the generality of Fig. 4. (a)(d) Local density of states of TiO$_2$ layers along $c$-axis // $P_S$. (b)(e) Ion positions near the top and bottom surface along $c$-axis; Green, blue and red filled circles show Sr, Ti and O ions, respectively. Tripods show nearest neighbor bonding that corresponds to the orientation of $P_S$ of unitcell. Encountering dipoles near the top surface, similar to BaTiO$_3$. Unlike BaTiO$_3$, half encountering dipoles at the bottom of (e). (c)(f) Ti-O and Sr-O displacement normalized by bulk values, polarization charge $-\Delta P$ normalized by a bulk $P_S$, and potential along $c$-axis.

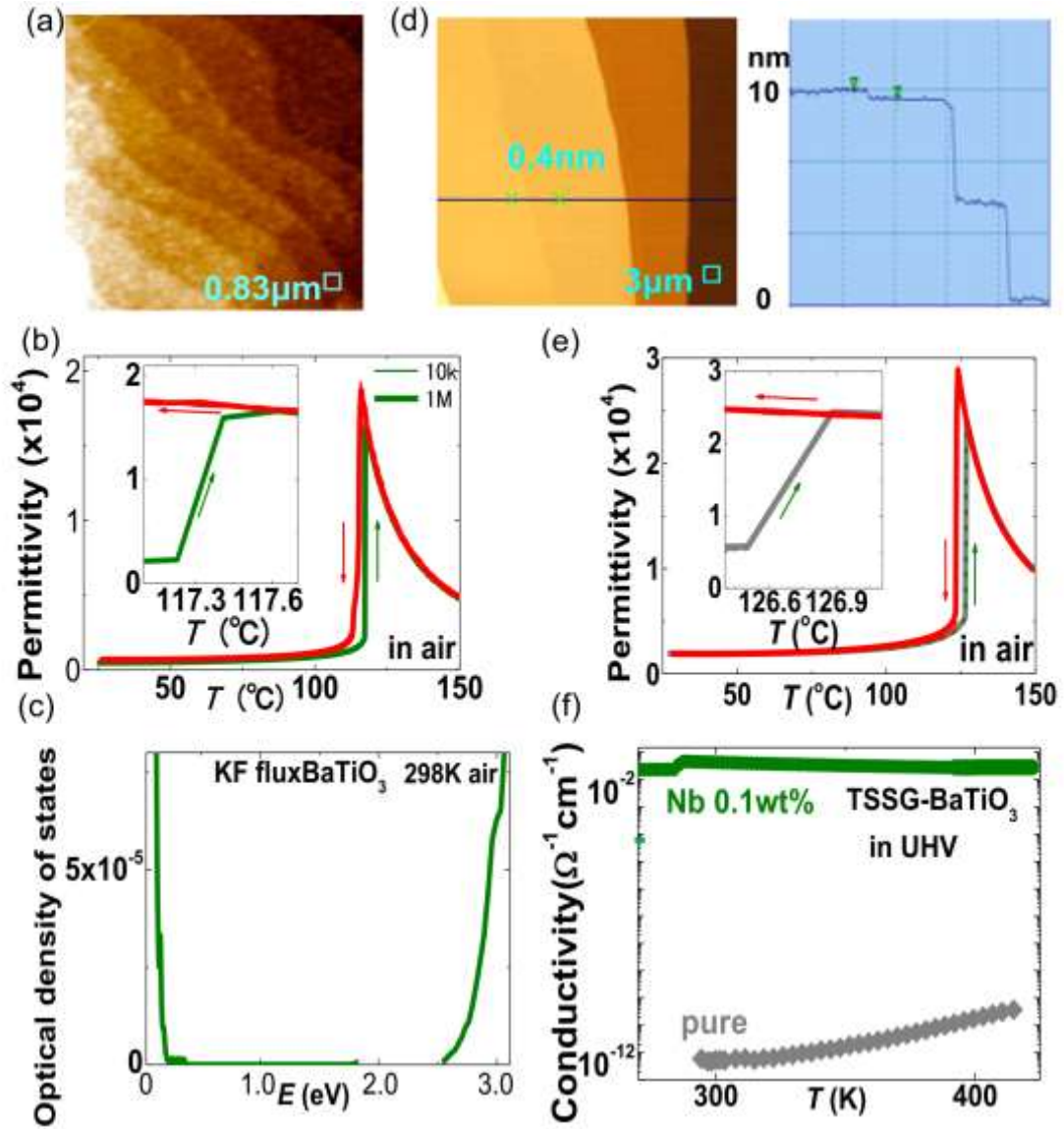

FIG. S15. Basic properties of samples: (a)-(c) KF-flux BaTiO$_3$ (d)-(f) TSSG BaTiO$_3$ used in Figs.1 and 2 (See also SM-1). (a) Surface after the atomic oxygen irradiation, preserving well-defined atomic steps of 1 unitcell. (b)(e) Permittivity in air showing sharp phase transitions. (c) Optical density of states ($dN_{eff}/dE/(m_{eff}/m_0)$(/eV/cell) in air. (d) Surface covered by single and multiple atomic steps achieved using only chemical etching and air-annealing without any mechanical polishing. This TSSG-BaTiO$_3$ crystal was etched by 25μm×2 to remove the surface layers possibly damaged by polishing. (f) High insulativity showing $10^{11}$ higher bulk resistivity than that of 0.1wt% Nb doped BaTiO$_3$.